\documentclass[usenatbib,fleqn,desactivate]{aa}

\pdfoutput=1

\usepackage{amsmath}
\usepackage{amssymb}
\usepackage{blindtext}
\usepackage{bm}
\usepackage{color}
\usepackage{graphicx}
\usepackage{hyperref}
\usepackage{lipsum}
\usepackage{longtable}
\usepackage{multicol}
\usepackage{multirow}
\usepackage{natbib}
\usepackage{txfonts}
\usepackage{wasysym}
\usepackage[capitalize,nameinlink]{cleveref}

\creflabelformat{equation}{#2#1#3}
\crefrangelabelformat{equation}{#3#1#4 to #5#2#6}

\hypersetup{colorlinks,
  linkcolor=red,
  filecolor=red,
  urlcolor=magenta,
  citecolor=blue}

\usepackage{array}
\newcolumntype{L}[1]{>{\raggedright\let\newline\\\arraybackslash\hspace{0pt}}m{#1}}
\newcolumntype{C}[1]{>{\centering\let\newline\\\arraybackslash\hspace{0pt}}m{#1}}
\newcolumntype{R}[1]{>{\raggedleft\let\newline\\\arraybackslash\hspace{0pt}}m{#1}}

\newcommand{\panel}[1]{\textit{#1}}
\newcommand{\leftpanel}{\panel{left}}
\newcommand{\rightpanel}{\panel{right}}
\newcommand{\Leftpanel}{\panel{Left}}
\newcommand{\Rightpanel}{\panel{Right}}
\newcommand{\midpanel}{\panel{middle}}

\newcommand{\toppanel}{\panel{top}}
\newcommand{\bottompanel}{\panel{bottom}}

\newcommand{\eagle}{EAGLE}
\newcommand{\hbt}{\textsc{hbt+}}

\newcommand{\msub}{m_\mathrm{sub}}

\newcommand{\mgas}{m_\mathrm{gas}}

\newcommand{\mstar}{m_{\star}}
\newcommand{\logmstar}{\log\mstar/\Msun}
\newcommand{\Msun}{\mathrm{M}_\odot}

\newcommand{\mtwo}{M_\mathrm{200m}}

\newcommand{\nbody}{$N$-body}
\newcommand{\percent}{\%}
\newcommand{\rtwo}{R_\mathrm{200m}}
\newcommand{\subfind}{\textsc{subfind}}
\newcommand{\tng}{IllustrisTNG}

\newcommand{\iacc}{i_\mathrm{acc}}
\newcommand{\icent}{i_\mathrm{cent}}
\newcommand{\iinfall}{i_\mathrm{infall}}
\newcommand{\isat}{i_\mathrm{sat}}
\newcommand{\imsub}{i_\mathrm{sub,max}}
\newcommand{\imstar}{i_\mathrm{\star,max}}
\newcommand{\imgas}{i_\mathrm{gas,max}}
\newcommand{\tacc}{t_\mathrm{acc}}
\newcommand{\tcent}{t_\mathrm{cent}}
\newcommand{\tinfall}{t_\mathrm{infall}}
\newcommand{\tmsub}{t_\mathrm{sub,max}}
\newcommand{\tmstar}{t_\mathrm{\star,max}}
\newcommand{\tmgas}{t_\mathrm{gas,max}}
\newcommand{\tsat}{t_\mathrm{sat}}
\newcommand{\zacc}{z_\mathrm{acc}}
\newcommand{\zcent}{z_\mathrm{cent}}
\newcommand{\zinfall}{z_\mathrm{infall}}

\newcommand{\zmstar}{z_\mathrm{\star,max}}

\newcommand{\zsat}{z_\mathrm{sat}}

\newcommand{\meangiven}[2]{\langle #1\mid #2 \rangle}
\newcommand{\scattergiven}[2]{\sigma(#1\mid #2)}

\begin{document}

\title{The history and mass content of cluster galaxies in the \eagle\ simulation}

\author{
    Crist\'obal Sif\'on\inst{1}
    \and
    Jiaxin Han\inst{2,3,4}
    }

\institute{
    Instituto de F\'isica, Pontificia Universidad Cat\'olica de Valpara\'iso, Casilla 4059, Valpara\'iso, Chile\\
    \email{cristobal.sifon@pucv.cl}
    \and
    Department of Astronomy, School of Physics and Astronomy, Shanghai Jiao Tong University, Shanghai, 200240, People’s Republic of China\\
    \email{jiaxin.han@sjtu.edu.cn}
    \and
    Key Laboratory for Particle Astrophysics and Cosmology (MOE), Shanghai 200240, China
    \and
    Shanghai Key Laboratory for Particle Physics and Cosmology, Shanghai 200240, China
    }

\date{Received XXX; accepted XXX}

\abstract
{}
{We explore the mass content of galaxies residing in galaxy clusters at $z=0$ in the \eagle\ cosmological hydrodynamical simulation. We also explore  the galaxies' mass build-up through cosmic time.}
{We used a galaxy catalogue generated with the \hbt\ algorithm, which identifies subhaloes consistently over time by tracking their dynamical evolution throughout the simulation.}
{The satellite subhalo-to-stellar mass relation (SHSMR) is well described by a double power law, becoming increasingly steeper with stellar mass. At stellar masses $9<\log\mstar/\Msun<10$, satellites have 20-25\percent\ the subhalo mass of central galaxies at fixed stellar mass. At high stellar masses, $\mstar>2\times10^{11}\,\Msun$, the satellite SHSMR is consistent with that of centrals. The satellite SHSMR decreases steeply for satellites closer to the cluster centre, even in projection, broadly consistent with recent weak lensing measurements.
The scatter in the satellite SHSMR is larger than that of central galaxies at all cluster masses and cluster-centric distances $R<R_\mathrm{200m}$. The SHSMR scatter decreases with stellar mass by about 12\percent\ over an order of magnitude, but this dependence can be explained by the mixing of infall times when binning by stellar mass.
By splitting satellites into direct and indirect infallers (those that fell into their current host as a central galaxy or as the satellite of an infalling group, respectively) we clearly show the impact of pre-processing separately on satellite galaxies' dark and stellar mass.
There is significant dark matter pre-processing; the most recent infallers into massive clusters ($\mtwo\gtrsim10^{14}\,\Msun$)  had already lost up to 50\percent\ of their dark matter by the time of infall, particularly if they fell in indirectly as satellites of another host.
On the contrary, on average, satellite galaxies are still gaining stellar mass at the time of infall and they do so for another 2 Gyr afterwards, although we see evidence of a slowing growth for indirect infallers. How much and for how long they continue to gain stellar mass depends primarily on the gas mass fraction available at infall.
Overall, pre- and post-processing  have similar impacts on the satellite SHSMR.
Finally, we provide a simple prescription to infer the mean mass loss experienced by satellites as a function of cluster-centric distance based on a comparison to central galaxies, convenient for observational weak lensing measurements.
}
{}

\maketitle

\section{Introduction}
\label{s:intro}

Over time, galaxies evolve and co-evolve as they are pulled together by gravity and merge into larger galaxies or form larger systems, such as groups and clusters of galaxies. Dynamical timescales in these objects are long, so a large fraction of the galaxies that fall into them survive as satellite galaxies. 
As this process unfolds, tidal interactions as well as hydrodynamic processes transfer mass from the satellites to the host system, changing the composition of a galaxy depending on its environment, and the dark matter haloes hosting these galaxies become subhaloes bound to the parent structure \citep[e.g.][]{Gunn1972,White1978,Springel2001}. Both gravitational and hydrodynamical interactions then unfold, transforming blue star-forming spiral galaxies into `red and dead' passively evolving ellipticals, giving rise to the well-known morphology-density relation \citep{Dressler1980} and a tight red sequence that is most prominent in massive galaxy clusters \citep{Gladders2000,Koester2007}. The clearest observational evidence of these interactions is found in `jellyfish' galaxies, in which radio observations reveal gas being removed from infalling galaxies, sometimes accompanied by trails of  star-forming regions observable in optical wavelengths \citep{Ebeling2014,Jaffe2018}.

A similarly dramatic experience is expected for the dark matter halo of a galaxy driven by tidal interactions. Dark matter is more loosely bound than stars are (and stars are heavily concentrated in the centres of galaxies), and is therefore more susceptible to tidal stripping \citep[e.g.][]{Chang2013,Smith2016}. Thus, as a galaxy spends more time within a larger halo it progressively and preferentially  loses  more dark matter; its total-to-stellar mass ratio decreases with time. Directly probing the dark matter content of galaxies requires gravitational lensing or spectroscopic measurements of stellar velocities within galaxies (i.e. rotation curves or velocity dispersions). Such measurements have provided tentative evidence of mass stripping 
by evaluating changes in galaxies' mass content or, in some cases, their mean sizes, with distance from the cluster centre as a proxy for time-since-infall. Early gravitational lensing analyses suggested that galaxies are significantly more concentrated the closer they are to a cluster centre, which was taken as evidence of mass loss through stripping \citep[e.g.][]{Limousin2005,Limousin2007,Natarajan2009}, although it is possible that these results were at least partly driven by the  modelling choices \citep[e.g.][]{PastorMira2011}.
More recently weak lensing analyses measuring the average total mass in subhaloes have found that the total-to-stellar mass ratio decreases with cluster-centric distance, as expected \citep{Sifon2015_lens,Sifon2018_Meneacs,Niemiec2017,Li2016,Kumar2022,Wang2023}, but a quantitative comparison between works is complicated by differing definitions of subhalo mass and density profiles \citep{Sifon2018_Meneacs}. Because we base our comparison to observational work on the aforementioned lensing studies, and because tidal stripping is expected to directly shape the subhalo-to-stellar mass ratio \citep{Han2016}, in this paper we use the term `mass segregation' to mean changes in the subhalo-to-stellar mass relation as done by those authors, although we note that this term is sometimes used to mean changes in the mean stellar or subhalo masses alone \citep[e.g.][]{Joshi2017,Kim2020}.

The impact of tidal effects on the dark matter (sub)haloes has been assessed to great extent in dark matter-only  simulations, which tend to agree that dark matter subhaloes can lose as much as 90\percent\ of their mass over   timescale of a few Gyrs after entering a massive halo \citep[e.g.][]{Rhee2017}. A significant fraction of subhaloes in \nbody\ simulations are completely disrupted because of these interactions, but most of this may be attributed to numerical noise \citep{Klypin1999_overmerging,Han2016,vdBoschOHB2018} or to the difficulty in accurately separating subhaloes from parent haloes \citep{Knebe2011,Han2018}, which complicates their interpretation.

When comparing to observations, \nbody\ simulations provide a valuable first approach, but are limited when performing detailed comparisons, as highlighted in several well-known discordances \citep[e.g.][]{Klypin1999_missing,Moore1999,Boylan-Kolchin2011,Jauzac2016}, which are largely alleviated when gas heating by supernovae and active galactic nuclei are incorporated into the simulations \citep[e.g.][]{Gao2004,Zolotov2012,Brooks2013,Bahe2021}.
Therefore, simulations that incorporate these hydrodynamical effects are required in order to inform our interpretation (be it astrophysical or cosmological) of observations in more detail. Such efforts are now possible thanks to high-resolution cosmological hydrodynamical simulations, which allow predictions of galaxy observables instead of only dark matter haloes \citep[e.g.][]{Vogelsberger2014,Schaye2015,Pillepich2018}.  Recent studies have made use of hydrodynamical simulations to establish more realistic expectations for the mass content of satellite galaxies \citep{Engler2021,Niemiec2022,Oneil2023}. 
Characterizing the relation between luminous and dark matter in satellite galaxies is also useful to allow the population of dark matter-only simulations with galaxies with realistic properties and evolution \citep[e.g.][]{Lacey2016,Jiang2021,Icaza-Lizaola2023}.

In this paper we explore the mass content of satellite galaxies using an improved galaxy catalogue from the largest hydrodynamical simulation within the Evolution and Assembly of Galaxies and their Environments (EAGLE) project \citep{Crain2015,Schaye2015}. The catalogue was produced with the \hbt\ algorithm \citep{Han2012,Han2018}, which natively tracks subhaloes consistently over time. We describe the subhalo-to-stellar mass relation at $z=0$ and how it has evolved over time, paying particular attention to the mass loss experienced by infalling galaxies and the build-up of the relation between total and stellar mass through time.

This paper is organized as follows. We describe the \eagle\ simulation and the galaxy and cluster samples used in \Cref{s:simulation}. In \Cref{s:present} we look at the present-day relation between total and stellar mass and compare to available observations. In \Cref{s:history} we summarize the history of cluster galaxies and their mass content. In \Cref{s:hsmr_evol} we discuss the implications for the subhalo-to-stellar mass relation over time, and in \Cref{s:massloss_obs} we discuss observational proxies of mass loss. We summarize our results in \Cref{s:conclusions}.

Throughout, we use $\log$ to refer to base 10-logarithm. We use the notations $\meangiven{X}{Y}$ and $\scattergiven{X}{Y}$ to represent the mean and scatter in $X$ at fixed $Y$, respectively.
Unless specified otherwise, masses and other properties refer to the present time, $z=0$.
Lookback times are in units of gigayears ago, expressed as Gya.

\section{Simulation data}\label{s:simulation}

We used the output of a simulation in the Evolution and Assembly of Galaxies and their Environment (\eagle) project \citep{Crain2015,Schaye2015}. \eagle\ is a suite of cosmological hydrodynamical simulations with varying box sizes, resolutions, and baryonic feedback prescriptions including supernova and AGN feedback. The simulation we use in this work is labelled \texttt{RefL0100N1504} and has a box size of $(100\,\mathrm{Mpc})^3$, with $1504^3$ particles and particle masses of $1.81\times10^6\,\Msun$ and $9.70\times10^6\,\Msun$ for baryons and dark matter, respectively. The simulation was run assuming a flat $\Lambda$CDM cosmology with $\Omega_\mathrm{m}=0.307$, $\Omega_\mathrm{b}=0.04825$, $\sigma_8=0.8288$, and $h=0.6777$. 

We used the subhalo catalogue produced with the upgraded Hierarchical Bound Tracing~\citep[\hbt,][]{Han2012,Han2018} post-processing of \eagle, in which subhaloes are found consistently across time by tracing their evolution as well as their instantaneous properties. This makes \hbt\ both more robust in identifying subhaloes and more stable in building their evolutionary histories. In particular, \hbt\ is robust against background obscuration \citep{Han2012}, which  prevents most subhalo finders from identifying subhaloes near the centre of the host \citep{Knebe2011,Muldrew2011}. By construction, it is also immune to the flip-flop problem when linking subhaloes across time \citep{Behroozi2015}. These advantages make \hbt\ an ideal choice for this work.\\

\subsection{Mass definitions and sample}

Throughout this work we speak of `subhaloes' as all dark matter haloes that inhabit a larger halo, including central subhaloes (the subhaloes associated with central galaxies); and of `galaxies' as the subset of subhaloes which contain at least one star particle. In practice, we limit our study to galaxies with stellar masses $\mstar\geq10^9\,\Msun$. All such subhaloes have at least 1,000 particles in total, at least 500 of which are stellar particles. Total masses of both subhaloes and galaxies are  defined as the total mass of all bound particles (i.e. the sum of all dark matter, gas, star, and black hole particles, defined in \hbt\ as \texttt{Mbound}), and stellar masses are similarly defined as the total mass in all bound star particles according to \hbt. Central and satellite subhaloes are hosted within a `host halo', whose total mass is defined as an overdensity mass $M_\mathrm{200m}$ with respect to the mean density of the Universe at the corresponding redshift, contributed by the masses of all the particles within the corresponding radius, $R_{\rm 200m}$. 

Although it is customary in analyses of numerical simulations to express masses in units of $h^{-1}\Msun$, throughout this work we take into account the value $h=0.6777$ assumed in the simulations and express masses in units of $\Msun$ alone.

We select all satellite galaxies with stellar masses $\mstar>10^9\,\Msun$ in galaxy clusters with $\mtwo>10^{13}\,\Msun$ at $z=0$, as calculated by \hbt. There are 216 clusters above this mass limit in \eagle, of which 68 have $\mtwo>5\times10^{13}\,\Msun$ and 12 have $\mtwo>10^{14}\Msun$; the most massive cluster in \eagle\ has a mass $\mtwo=6.1\times10^{14}\,\Msun$. These 216 clusters host a total of 3,740 satellite galaxies above our stellar mass limit of $10^9\,\Msun$, of which 1,494 reside in $\mtwo>10^{14}\,\Msun$ clusters.

In some cases it is useful to compare the properties of satellites to those of centrals of the same stellar mass. With the stellar mass cut of $9<\log\mstar/\Msun<11$, there are in total 7,205 centrals in the simulation box at $z=0$. These centrals reside mostly in lower mass haloes than clusters, with masses typically in the range $11<\log\mtwo<12.5$.

\section{Present-day mass content of cluster galaxies in \eagle}\label{s:present}

\begin{figure}
    \centering
    \includegraphics[width=0.95\linewidth]{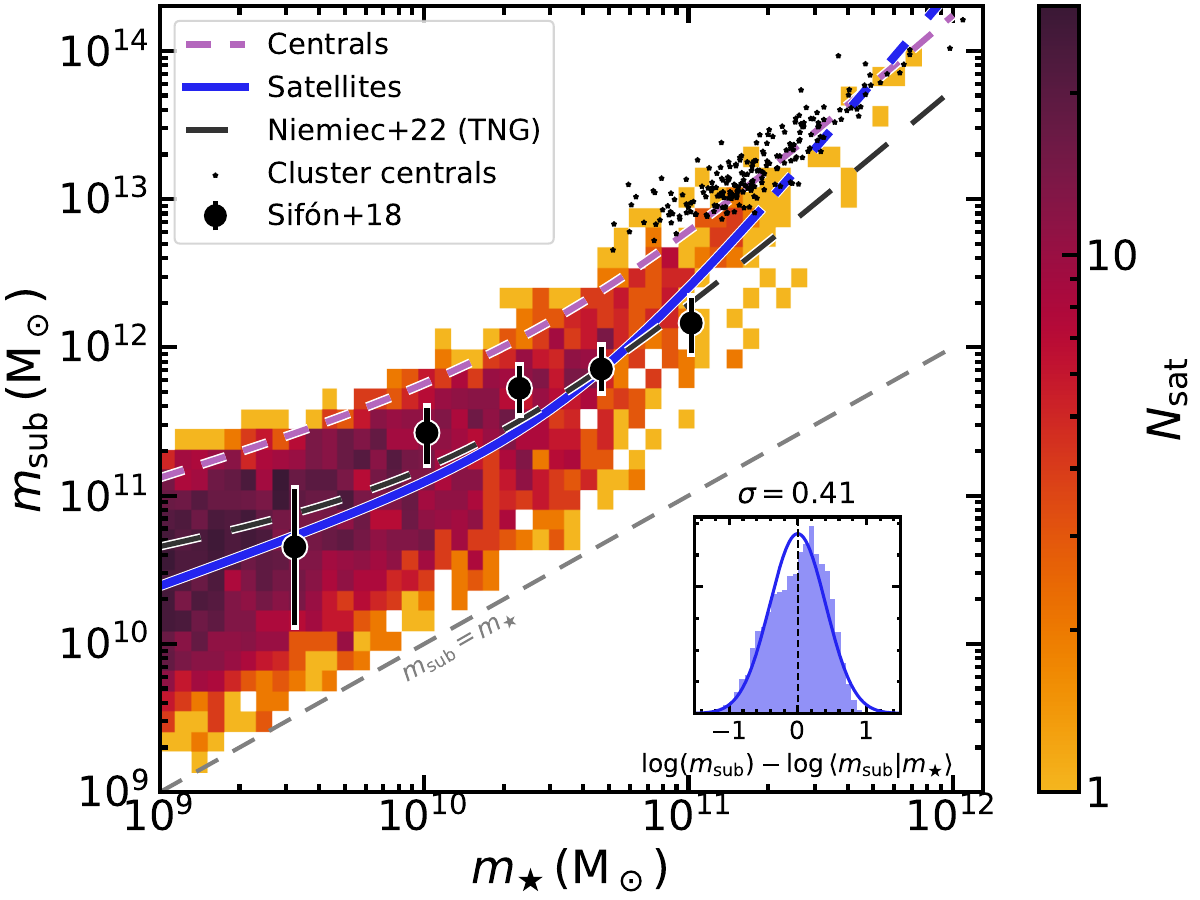}
    \caption{Subhalo-to-stellar mass relation (SHSMR) for satellite galaxies in galaxy clusters with masses $\mtwo>10^{13}\,\Msun$ in the EAGLE simulation. The coloured background shows the occupation of the $\msub-\mstar$ plane, while the thick blue line shows the best-fitting \Cref{eq:hsmr}, which has been fit only over the range in which the blue line is solid. The dashed magenta line shows the corresponding fit to central galaxies while black dots show the centrals in the $\mtwo>10^{13}\,\Msun$ clusters included in this work. The dashed black line black line shows the satellite SHSMR in \tng\ \citep{Niemiec2022} and large circles with error bars show the weak lensing measurements of \citet{Sifon2018_Meneacs}.
    The inset shows the logarithmic scatter around the mean relation, which can be approximated by a lognormal distribution with standard deviation $\sigma=0.41$.
    }
    \label{f:hsmr}
\end{figure}

We begin by studying the relation between satellite galaxies' subhalo and stellar masses at $z=0$. In \Cref{s:history,s:hsmr_evol} we then study the evolution of satellites and their mass content over cosmic time, with the aim of understanding the physical origin of the present-day SHSMR.

\begin{figure}
  \centerline{
    \includegraphics[width=\linewidth]{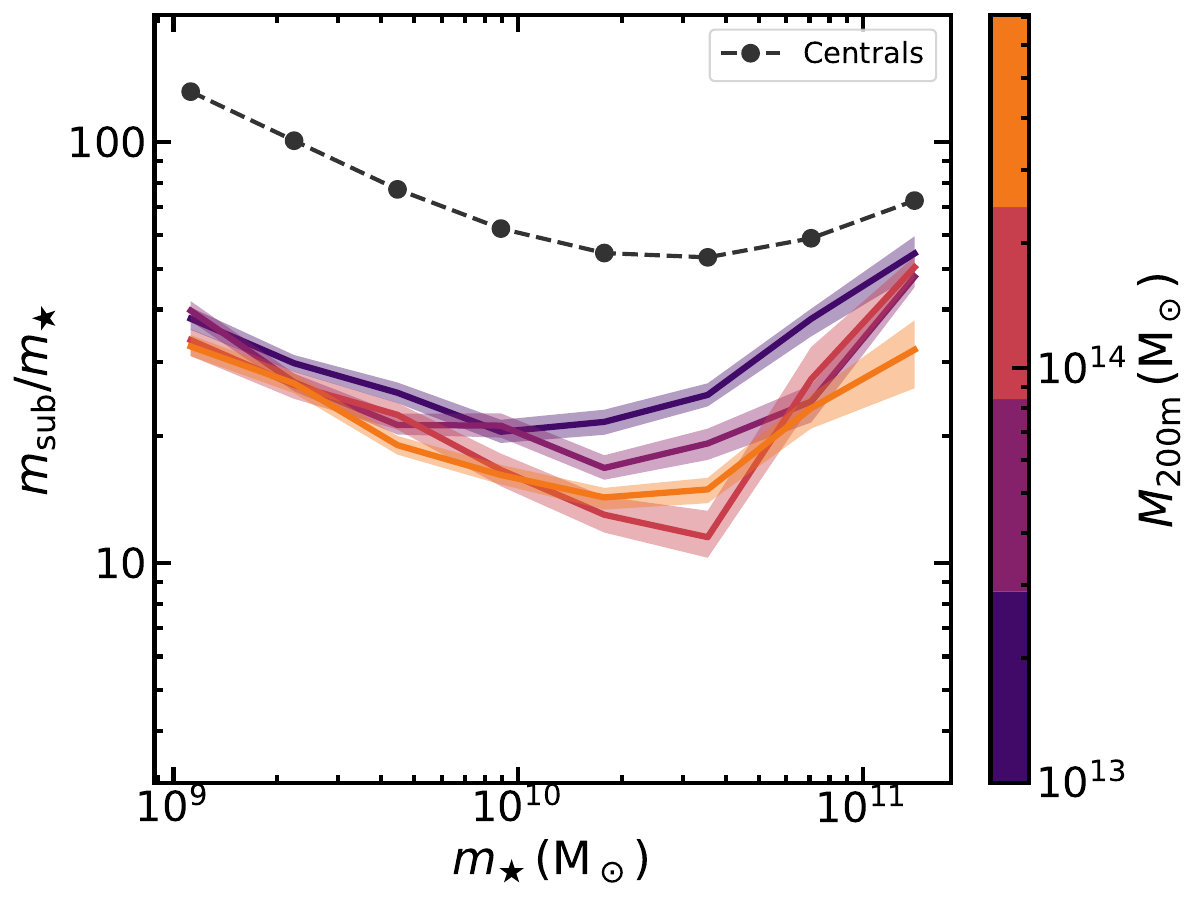}
    }
  \caption{Dependence of the satellite subhalo-to-stellar mass relation on host cluster mass, $\mtwo$, as described by the colour bar, where jumps in colour mark $\mtwo$ bin edges. The shaded regions are standard errors on the means. The black dashed line with circles shows the mean SHSMR of central galaxies.
  }
  \label{f:hsmr_m200}
\end{figure}

\begin{figure*}
    \centering
    \includegraphics[width=0.48\linewidth]{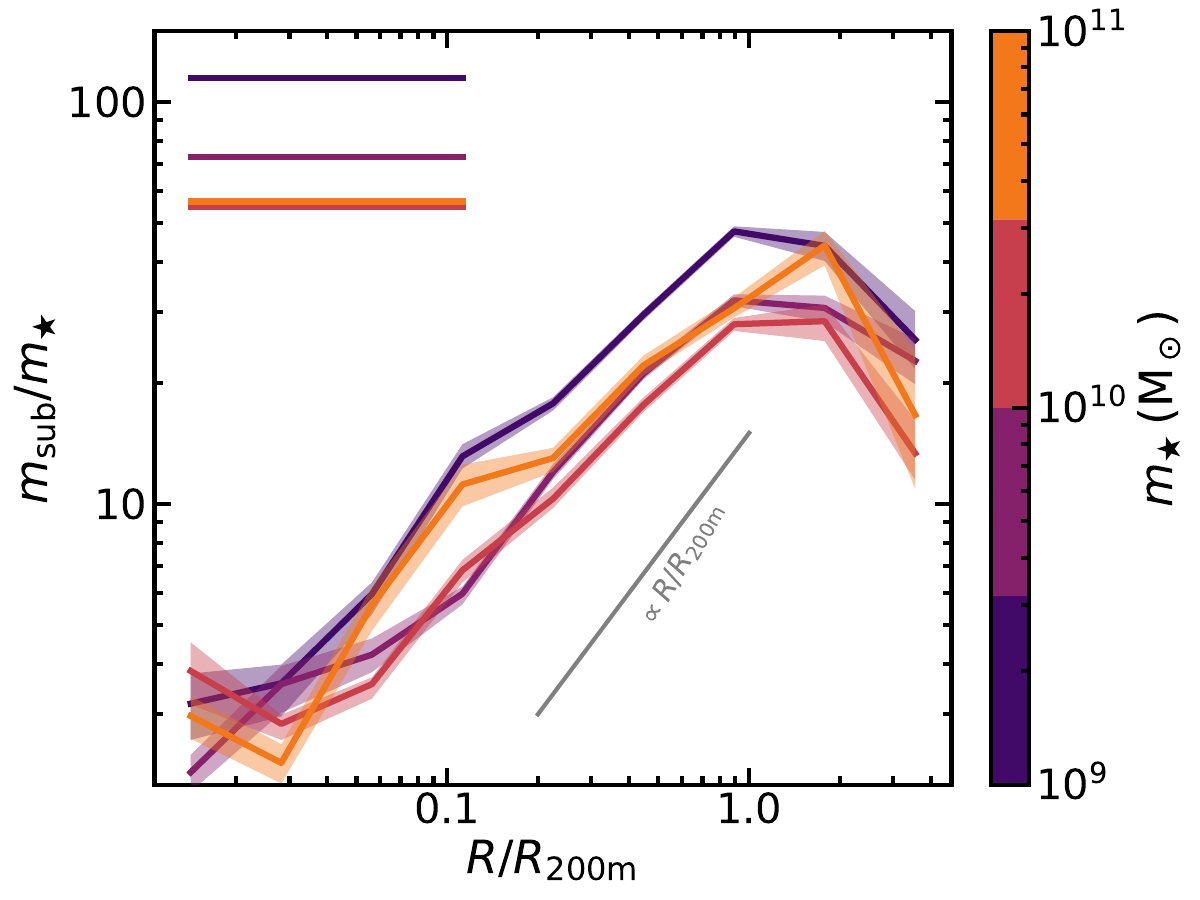}
    \includegraphics[width=0.48\linewidth]{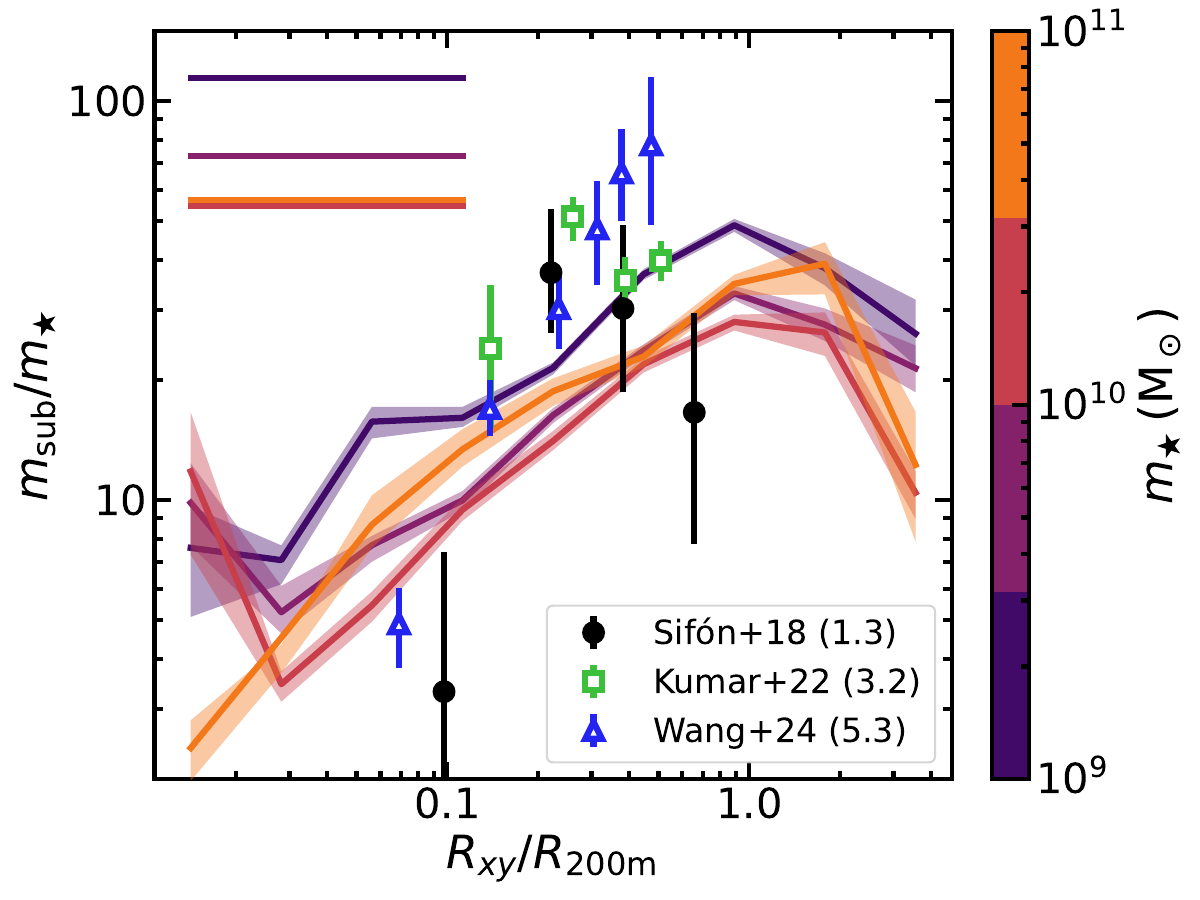}
    \caption{Total-to-stellar mass ratio (SHSMR) as a function of cluster-centric distance normalized by cluster size, $R_\mathrm{200m}$, binned by stellar mass. Horizontal lines in the top left show the mean SHSMR for centrals in each stellar mass bin. \Leftpanel: 3D distances. The straight grey line shows a linear dependence with cluster-centric distance, for reference. \Rightpanel: Distances projected along the $z$ direction. 
    Weak lensing measurements from \citet{Sifon2018_Meneacs}, \citet{Kumar2022}, and \citet{Wang2023} are shown with black circles, green squares, and blue triangles, respectively. Numbers in parentheses correspond to the mean stellar mass in each study, in units of $10^{10}\,\Msun$.
    }
    \label{f:mratio_dist}
\end{figure*}

\subsection{Subhalo-to-stellar mass relation}
\label{s:shsmr}

\Cref{f:hsmr} shows the joint distribution of subhalo and stellar masses for satellite galaxies in massive clusters in \eagle, as well as the subhalo-to-stellar mass relation (SHSMR),  $\meangiven{\msub}{\mstar}$, for central and satellite galaxies. Following \cite{Niemiec2022} we fit the satellite and central SHSMRs separately using double power laws of the form
\begin{equation}\label{eq:hsmr}
    \meangiven{\msub}{\mstar} 
    = 2N\left[\left(\frac{\mstar}{m_1}\right)^{-\beta} 
            + \left(\frac{\mstar}{m_1}\right)^\gamma\right] \mstar.
\end{equation}

\begin{table}
    \centering
    \caption{Best-fit subhalo-to-stellar mass relation parameters from \Cref{eq:hsmr} for central and satellite galaxies.}
    \label{t:hsmr}
    \begin{tabular}{c r@{$\,\pm\,$}l r@{$\,\pm\,$}l}
    \hline
     Parameter & \multicolumn{2}{c}{Centrals} & \multicolumn{2}{c}{Satellites}\\[0.5ex]
     \hline
     $\log m_1/\Msun$ & $10.54$ & $0.05$ & $10.56$ & $0.09$\\
     $\beta$ & $0.46$ & $0.02$ & $0.36$ & $0.04$\\
     $\gamma$ & $0.57$ & $0.03$ & $1.11$ & $0.16$\\
     $N$ & $12.55$ & $0.11$ & $3.39$ & $0.19$ \\[0.5ex]
    \hline
    \end{tabular}
\end{table}

Best-fit parameters for centrals and satellites are listed in \Cref{t:hsmr}. The ratio of satellite-to-central SHSMR increases from $0.19\pm0.04$ at $\mstar=10^9\,\Msun$ to consistent with unity at $\mstar>2\times10^{11}\,\Msun$.
As shown in \Cref{f:hsmr}, over the range $9<\logmstar\lesssim11$ the EAGLE SHSMR agrees with the SHSMR of satellite cluster galaxies in the \tng\ simulation derived by \cite{Niemiec2022}. At larger stellar masses the EAGLE SHSMR tends to lie above that from IllustrisTNG. This could be largely explained by the different subhalo finders used. The most massive subhaloes usually have their outer part well mixed with their host haloes, making it difficult to separate the two using only a single snapshot. In \tng, subhaloes were identified using the (instantaneous) configuration-space subhalo finder \subfind, which has difficulty determining the masses of subhaloes deep into their host haloes \citep[e.g.][]{Knebe2011,Onions2012,Han2018}.
As \hbt\ uses temporal information when identifying subhaloes, it is able to recover a much more complete list of member particles for these massive subhaloes. Most notably, the SHSMR for satellites at the highest mass end approaches that for central galaxies according to \hbt. 
This is consistent with the discussions in \citet{Han2018} that the most massive subhaloes experience weaker tidal stripping and stronger dynamical friction, so that the surviving massive ones are not much different from centrals.
Nevertheless, to draw conclusions that are not sensitive to this halo-finding systematic, in the remainder of this work we study satellite galaxies over the range $9<\logmstar<11$. As we discuss next, this is also the range over which there are observational measurements to compare with.

We also compare the EAGLE SHSMR with galaxy-galaxy lensing measurements of satellites galaxies in massive ($M_\mathrm{200m}\gtrsim10^{14}\,\Msun$) clusters \citep{Sifon2018_Meneacs}, which is consistent with measurements at lower cluster masses \citep[e.g.][]{Dvornik2020}, in \Cref{f:hsmr}. The stellar mass range probed by \cite{Sifon2018_Meneacs} stops right where EAGLE predicts a turn in slope so that the weak lensing measurements are not able to assess the double power-law nature of the SHSMR, but over the range probed the EAGLE SHSMR agrees with the observational measurements.
\Cref{f:hsmr_m200} shows that the SHSMR\footnote{In order to reduce the dynamical range, we represent the SHSMR as $\meangiven{\msub/\mstar}{\mstar}$ hereafter.} depends weakly on cluster mass.
At low stellar masses, $\msub\lesssim10^{10}\,\Msun$, there is a constant $\sim$10\percent\ difference in $\msub/\mstar$ over an order of magnitude in cluster mass. 
At higher stellar masses, $10^{10}\lesssim\mstar/\Msun\lesssim10^{11}$, galaxies have approximately twice as much total mass at fixed stellar mass in $\mtwo\sim10^{13}\,\Msun$ than in $\mtwo\sim10^{14}\,\Msun$ clusters, although this trend seems to stall for larger cluster masses. That this difference manifests itself only for more massive satellites can be generally ascribed to their shorter dynamical timescales. The trend with cluster mass, on the other hand, can be due to two effects, noting that---as discussed later---dark matter is more susceptible to mass loss than stellar mass. Firstly, tidal stripping is expected to be stronger in more massive clusters. Secondly, galaxies in more massive clusters are expected to have been more strongly pre-processed, such that they enter massive clusters already with a reduced dark matter fraction \citep{Bahe2019}. We return to this in \Cref{s:hsmr_evol}.

\subsection{Subhalo mass segregation}
\label{s:segregation}

We now turn to the spatial trends we may expect to see in observational data.
\cite{Han2016} studied the spatial distribution of subhaloes in dark matter-only simulations, finding that they closely follow the dark matter particles. This implies that dynamical friction does not play an important role, at least for low mass subhaloes or beyond $R>0.02\rtwo$. They showed that differences in the distribution of subhaloes and dark matter particles in the literature are due not to dynamical friction but to the selection of subhaloes based on present-day subhalo mass instead of mass at accretion \citep[see also][]{vandenBosch2016,He2023}. Our selection is based on present-day stellar mass, which is expected to be less affected by these selection biases. In addition, any such biases will also be present in observations, for which we aim to provide a reference point.

We show the present-day mass segregation of EAGLE galaxies, expressed as the mean ratio $\msub/\mstar$ as a function of cluster-centric distance at fixed stellar mass, in the left panel of \Cref{f:mratio_dist}. Over the range $0.05<R/\rtwo<1$, this ratio is well described by a power law, $\msub/\mstar\propto (R/\rtwo)^{0.7}$.
At smaller radii the apparent segregation diminishes as a consequence of numerical resolution, which also reduces the scatter in the SHSMR (discussed in \Cref{s:hsmr_scatter}, see \Cref{f:scatter_r}). At larger radii, $R>\rtwo$, we see a turnover in the SHSMR, which can be attributed to the presence of galaxies that already reached pericentre and whose apocentre lies outside $\rtwo$ (i.e. these galaxies are currently far from the cluster centre but have already suffered significant mass loss). We return to these points in \Cref{s:hsmr_evol}.

The cluster-centric distance dependence of $\langle\msub/\mstar\rangle$ is similar to that of the ratio between present-day and maximum subhalo mass, $\msub^\mathrm{max}$, commonly measured for dark matter subhaloes in \nbody\ simulations. We find $\langle\msub/\msub^\mathrm{max}\rangle\propto(R/\rtwo)^{0.6}$ in \eagle, and the same slope was found by \cite{vandenBosch2016} in a number of \nbody\ simulations.\footnote{This value is not reported explicitly by \cite{vandenBosch2016} but can be extracted from the lower-left panel of their Figure 3.} This similarity between the dependence of $\msub/\mstar$ and $\msub^\mathrm{max}$ with cluster-centric distance is a consequence of the resilience of stellar mass to external forces, as discussed in more detail in later sections. This slope is different from that found by \cite{Han2016} for satellites in massive haloes \citep[see also][]{He2023}. However, we point out that \cite{Han2016} treated selection effects carefully in order to extract the parent subhalo population while we report the slope at face value in an attempt to reproduce the observational situation. Furthermore, because we are mostly concerned with a comparison with observations, we report the mean values, $\langle\msub/\msub^\mathrm{max}\rangle$; instead, since the goal of \cite{Han2016} was to understand the underlying subhalo population they report the median in logarithmic space, $\mathrm{median}[\log(\msub/\msub^\mathrm{max})]$. Using the same statistic as \cite{Han2016} we find a slope of 0.8, closer to (but still different than) the latter authors. The remaining difference with the slope of 1.0 from \cite{Han2016} may be due in part to selection effects and in part to satellite galaxies being somewhat more resistant to tidal stripping than dark matter subhaloes because of the increased central density.

Recently, large optical surveys with high image quality have allowed observational measurements of the dependence of the subhalo-to-stellar mass ratio with (projected) cluster-centric distance, i.e. subhalo mass segregation, using weak gravitational lensing. Although a detailed comparison between these results can be affected by modelling and analysis differences, overall these results do suggest mild segregation \citep{Sifon2015_lens,Sifon2018_Meneacs,Li2016,Niemiec2017,Kumar2022,Wang2023}.
In order to allow a comparison with observational work, we also show the dependence with projected distance, which we refer to as $R_{xy}$,\footnote{Here and throughout we use projected distances measured along the $x-y$ plane. We have verified that using the $x-z$ or $y-z$ planes produces statistically identical results, as expected.} in the right panel of \Cref{f:mratio_dist}. The decrease in $\msub/\mstar$ is still present but is reduced due to projection effects---to roughly a factor 2 instead of 10 between $R_{xy}=\rtwo$ and $R_{xy}=0.05\rtwo$ at fixed stellar mass. The segregation observed in \eagle\ is consistent with the results of both \cite{Sifon2018_Meneacs} and \cite{Kumar2022}, although the measurements of \cite{Kumar2022} tend to lie above the EAGLE relations, while \cite{Wang2023} found stronger segregation than seen in EAGLE. As discussed by \cite{Sifon2018_Meneacs}, this could be due to differing definitions of subhalo mass as well as stellar mass. An additional complication in the comparison to observations is that our analysis is entirely free of interlopers, i.e. galaxies mistakenly identified as cluster members. Because of the steepness of the dependence with cluster-centric distance and because central galaxies have significantly higher subhalo-to-stellar mass ratios than satellites, this contamination will tend to increase the inferred mean mass. We leave a detailed assessment of interlopers for future work.

We note that some works have studied the segregation of mean stellar mass alone rather than the subhalo-to-stellar mass ratio. For instance, \cite{Roberts2015} found that segregation of stellar mass decreases with increasing cluster mass, and that low-mass galaxies are more strongly segregated. At cluster masses typical of our sample, they find a slope of roughly $-0.02$ dex, i.e. mean stellar mass increasing slightly towards the cluster centre. While we do not attempt to match both samples in detail, we measure an overall slope of $\log\mstar\propto-0.03(R/\rtwo)$, consistent with \cite{Roberts2015}.

The strength of the segregation does not depend on stellar mass (i.e. all curves in \Cref{f:mratio_dist} have the same slope). This is expected if the trend is due mostly to tidal stripping, since stellar mass is more centrally concentrated than dark matter and is therefore the component least affected by external tidal forces \citep{Chang2013,Smith2016}, although the normalization changes following the SHSMR in \Cref{f:hsmr_m200}.
This latter dependence means that observationally one must ensure galaxies at different cluster-centric radii have comparable stellar mass distributions. In particular, if the mean stellar mass increases with radius by roughly half a dex or more then we would not expect to observe any segregation. However this is not the case for any of the recent weak lensing measurements, where the change is at most a factor 2, i.e. less than the bin size in \Cref{f:mratio_dist}. We would therefore expect the slopes seen in observations  to be comparable with those in the right panel of \Cref{f:mratio_dist}. Although there is considerable point-to-point variation, there is overall agreement between observations and simulations.

We also show in \Cref{f:mratio_dist} the mean $\msub/\mstar$ of central galaxies in each stellar mass bin (cf.\ \Cref{f:hsmr_m200}). Satellites have lower SHSMR than centrals at all radii. As discussed in \Cref{s:hsmr_evol}, this is a consequence of pre-processing, i.e. the influence of lower-mass host haloes on the galaxies' mass content before infall to the present host, which means satellites have lower masses than centrals even outside $\rtwo$.

\begin{figure}
    \centering
    \includegraphics[width=\linewidth]{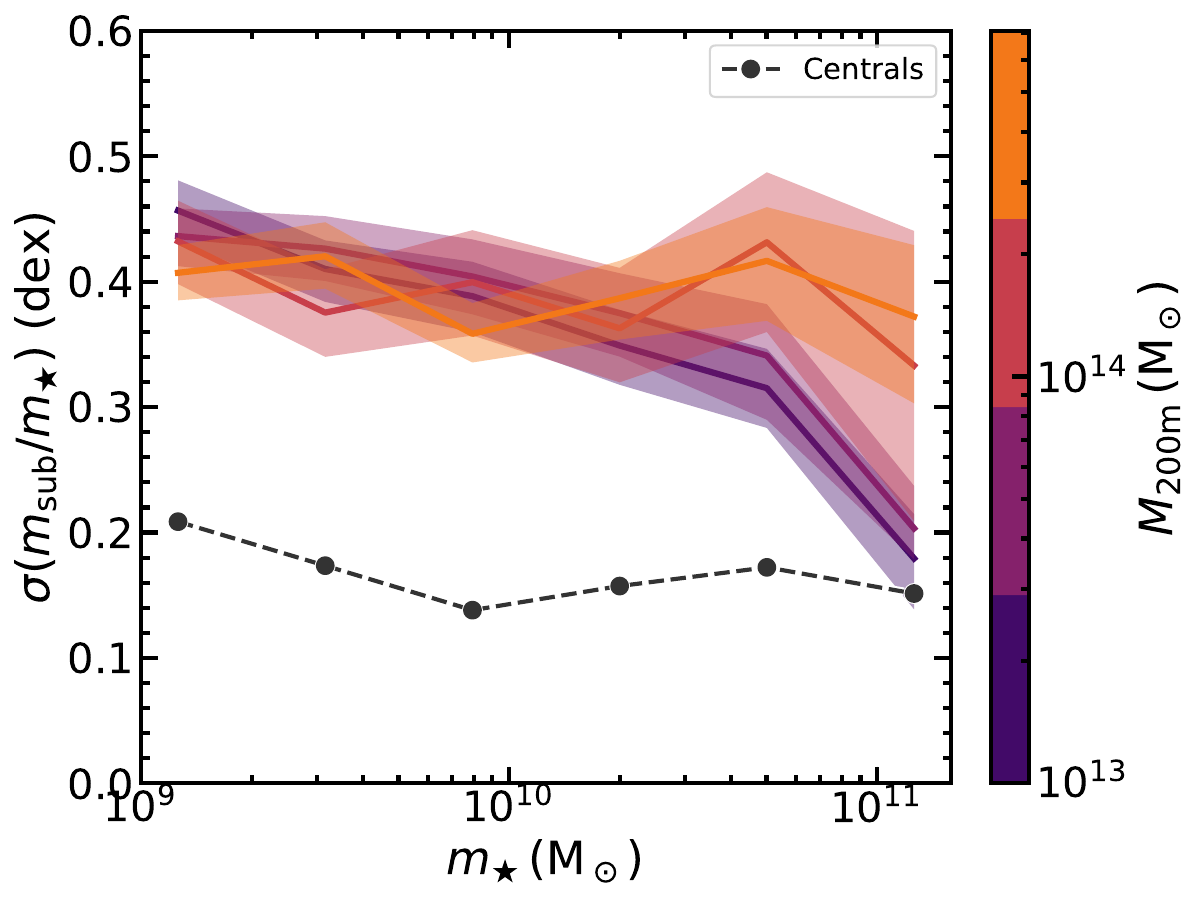}
    \caption{Dependence of the logarithmic scatter in the present-day SHSMR on host halo mass. The dashed black line with circles shows the scatter of central galaxies, for comparison.}
    \label{f:scatter_mhost}
\end{figure}

\subsection{The scatter in the relation between total and stellar masses in cluster galaxies}
\label{s:hsmr_scatter}

\begin{figure}
 \centering
 \includegraphics[width=\linewidth]{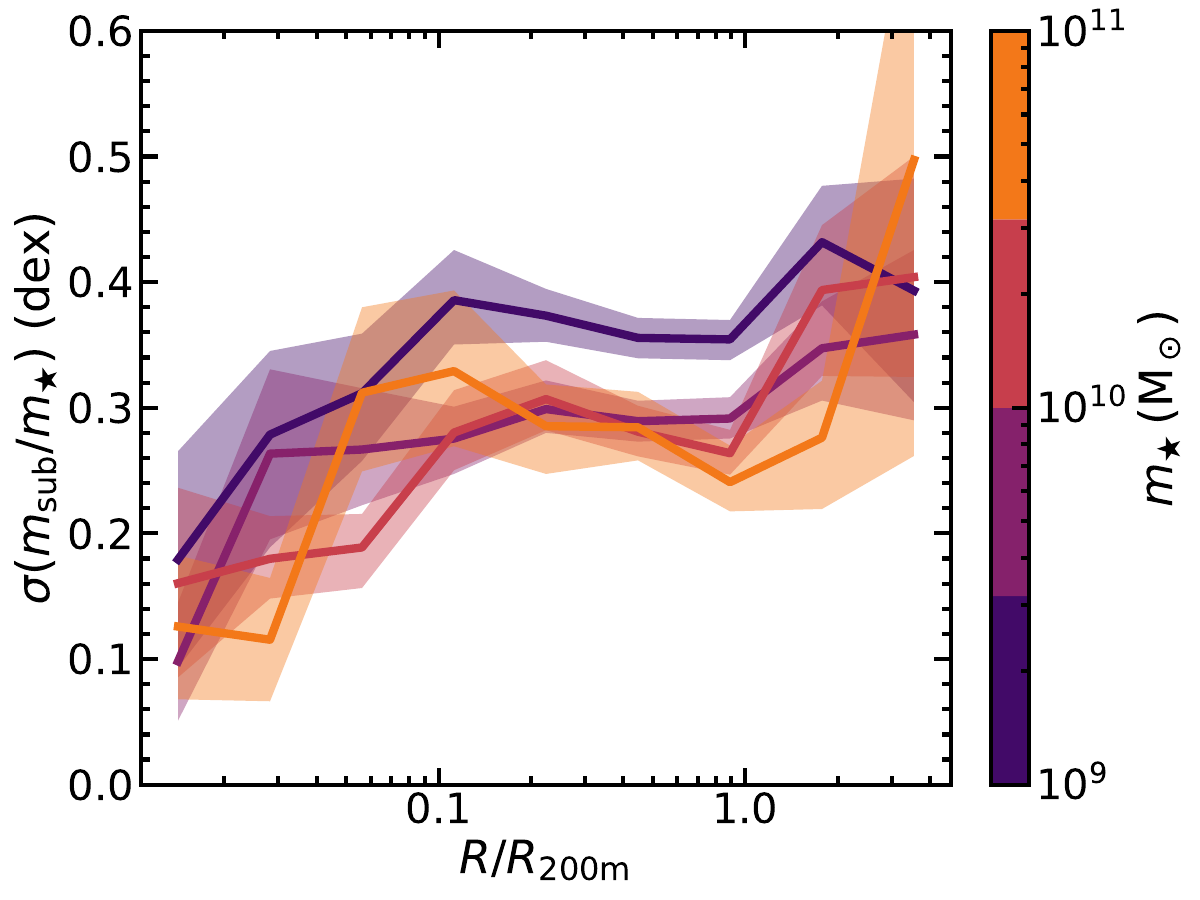}
 \caption{Dependence of intrinsic scatter in the SHSMR on three-dimensional cluster-centric distances, for satellite galaxies of different stellar masses.}
 \label{f:scatter_r}
\end{figure}

Another important aspect of the relation between total and stellar mass is the scatter in the relation, as it can tell us about the variety of phenomena that affect said relation. For instance, \cite{Niemiec2022} found that the scatter is mostly correlated with distance to the centre and the compactness of galaxies (in addition to properties specifying the history of the subhalo, which we explore in \Cref{s:hsmr_evol}), but it is unrelated to the properties of the host cluster. 

As shown in \Cref{f:scatter_mhost}, the scatter in the present-day satellite SHSMR tends to decrease slightly with stellar mass, and this behaviour shows no significant dependence on cluster mass. Considering all subhaloes, the scatter is $0.43\pm0.01$ dex at $\mstar=10^9\,\Msun$, while at $\mstar=10^{11}\,\Msun$ the scatter is $0.36\pm0.04$ dex, i.e. a decrease of $(17\pm11)$\percent\ over two orders of magnitude in stellar mass. This is statistically consistent with \cite{Niemiec2022} who find no dependence of SHSMR scatter with stellar mass. The scatter in the satellite SHSMR is larger than the central SHSMR, which also decreases slowly with stellar mass, from 0.20 dex at $\mstar=10^9\,\Msun$ to roughly 0.15 dex at $\mstar=10^{11}\,\Msun$. 

On the other hand, there is a slight but noticeable dependence of the SHSMR scatter on cluster-centric distance (\Cref{f:scatter_r}), with a scatter approaching that of centrals at the closest separations, $R\lesssim0.05\rtwo$ (70 kpc for a $\mtwo=10^{14}\,\Msun$ cluster at $z=0$). The scatter increases slowly outwards from $0.20\pm0.03$ dex at $R<0.02\rtwo$ to $0.43\pm0.02$ dex at $R>\rtwo$, when averaging over all stellar masses. As with $\langle\msub/\mstar\rangle$, this reduction in scatter may be due in part to the disruption of subhaloes, be it artificial or physical, whereby only the most massive end of the subhalo population survives very close to the cluster centre.

\Cref{f:scatter_r} shows the dependence of scatter on three-dimensional radius. There is no such trend in projection---there is even a hint of an inverted trend, with intrinsic scatter decreasing up to 20\percent\ from the innermost regions to just inside $\rtwo$, due to the mixing of satellites at different physical distances. Beyond $R_{xy}=\rtwo$, the intrinsic scatter increases again to $\approx0.40$ dex for all stellar masses.

\begin{figure*}
    \includegraphics[width=\linewidth]{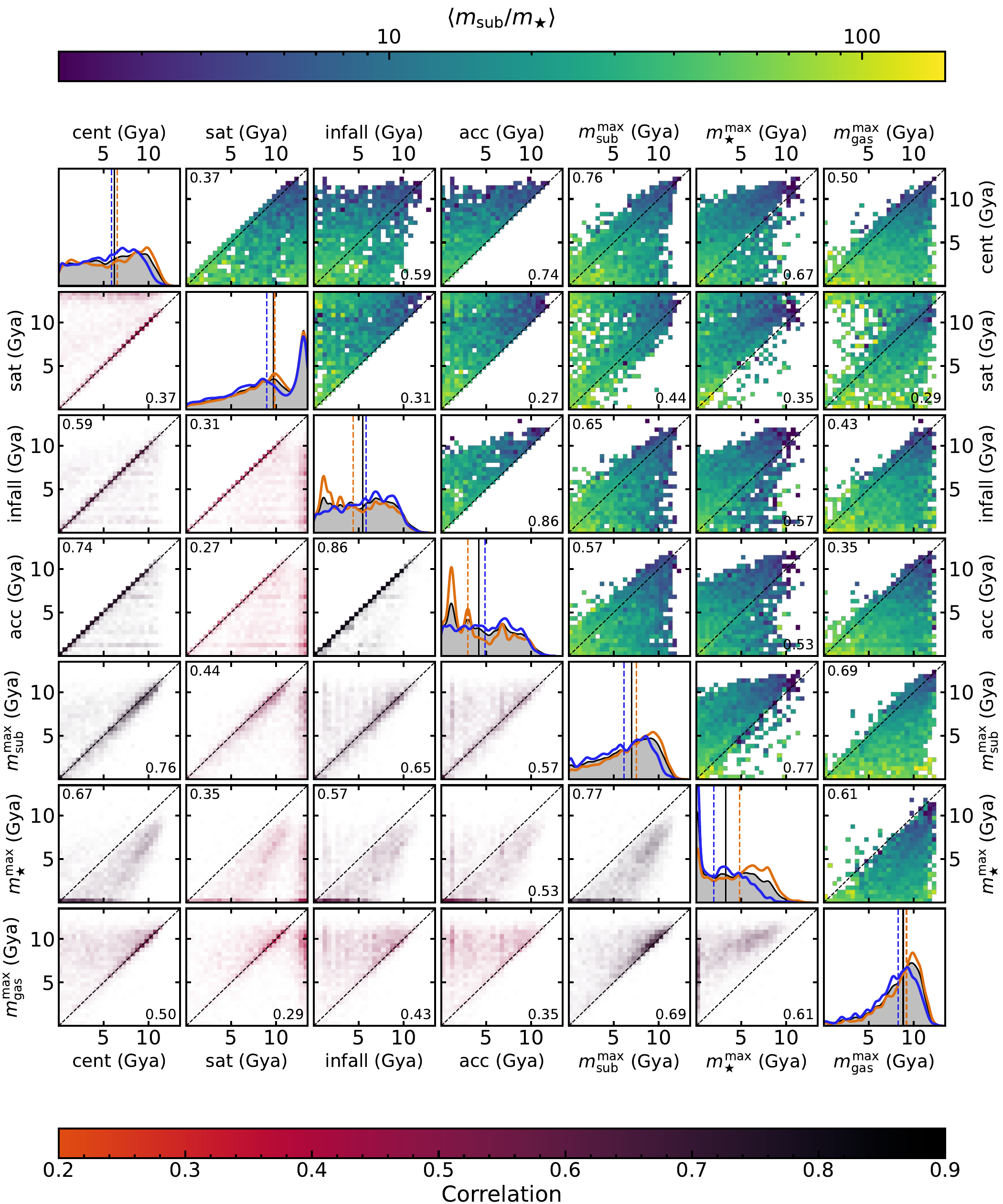}
    \caption{Correlation between the different lookback times listed in \Cref{t:times} (in units of   Gya).
    The \emph{diagonal panels} show kernel density estimates of each lookback time distribution, using a Gaussian kernel with $\sigma=0.1\,\mathrm{Gyr}$, and the median marked by a blue line. The orange and blue curves and dashed vertical lines show distributions and medians for satellites in $\mtwo>10^{14}\,\Msun$ and $10^{13}\,\Msun<\mtwo<5\times10^{13}\,\Msun$. Each off-diagonal panel in the \emph{lower triangle} shows the corresponding two-dimensional occupation; the colour scale runs from white (zero) to a colour given by the Spearman correlation coefficient between the two times, shown in the corner of each panel, and coded in the bottom colour bar. The brightness scaling is consistent across panels,
    with the darkest cells containing 2.5\percent\ of the population or more. Each panel in the \emph{upper triangle} shows the logarithm of the mean subhalo-to-stellar mass ratio at $z=0$ in each cell, as coded in the upper colour bar. The numbers in the corners of each panel  correspond to the correlation coefficients.
    }
    \label{f:times}
\end{figure*}

\section{A brief history of satellite galaxies}
\label{s:history}

Having established the statistical properties of cluster satellites at present, we begin this section by reviewing the history of the satellite galaxies they host, which provides some context for the results of the following subsections (and the rest of the paper) where we explore how the mass content of satellite galaxies changes over cosmic time and what has driven these changes.

\subsection{Milestones in the life of a satellite galaxy}
\label{s:milestones}

We identify the following key moments\footnote{For each of these moments we use $i$ to refer to the snapshot number (increasing with cosmic time), $t$ to refer to the time, and $z$ to the redshift, depending on the context.} in the life of a satellite galaxy: after being formed, we label $\icent$ the last snapshot in the simulation when a galaxy was labelled as central and $\isat$ the first snapshot it was labelled a satellite of any host, independent of the host's mass. We also label the first time a galaxy is identified as a satellite of its current host as $\iinfall$, and the first snapshot after which it never leaves its current host again as $\iacc$. By definition, then, $\zsat\geq\{\zcent,\zinfall\}\geq\zacc$. Additionally, we register the snapshots where each subhalo reached maximum total mass ($\imsub$), stellar mass ($\imstar$), and gas mass ($\imgas$). Throughout, we use ``infall'' to refer to infall to the $z=0$ host specifically.
The times $\tinfall$ and $\tacc$ are sometimes referred to as first and last infall, and are analogous to $t_\mathrm{branch}$ and $t_\mathrm{main}$ in \cite{Bahe2019}.\footnote{\cite{Bahe2019} defined $t_\mathrm{branch}$ and $t_\mathrm{main}$ as the middle of the relevant snapshot interval, whereas we used the specified snapshot time. This introduces typically a $\approx30$ Myr difference between those authors' and our definitions at $z<1$ and differences much smaller at earlier times; these are insignificant for our purposes.} $\tinfall$ is also similar to the residence time defined by \cite{Oneil2023}, except they defined it as the time since the subhalo first crossed $\rtwo$, while we define it based on whether it is gravitationally bound as calculated by \hbt.

\begin{table}
    \centering
    \caption{
    Statistics for present-day satellite galaxies' characteristic lookback times
    in Gigayears ago (Gya). Brackets show redshifts corresponding to the quoted lookback times. Times are, from top to bottom: \textbf{sat}: first time as a satellite (see text for caveats), \textbf{cent}: last time as a central, \textbf{infall}: first time labelled as a satellite of current host, \textbf{acc}: definitively labelled as a satellite of current host. We also list the times of maximum stellar, subhalo (i.e. total), and gas masses.
    }
    \label{t:times}
    \begin{tabular}{l|cc@{ -- }lr@{ -- }l}
    \hline\hline
        \multirow{2}{*}{Event} & \multicolumn{5}{c}{Lookback times (Gya)}\\
        {} &   Median    & \multicolumn{2}{c}{50\% range} & \multicolumn2c{90\% range} \\[0.5ex]
    \hline
        sat    & 9.4 [1.46] & 6.0 & 12.7 & 1.8 & 13.2 \\
        cent   & 5.3 [0.52] & 2.8 &  7.9 & 0.4 &  9.6 \\ 
        infall & 5.1 [0.48] & 2.7 &  7.5 & 0.7 &  9.0 \\
        acc    & 4.0 [0.35] & 1.8 &  6.7 & 0.2 &  8.5 \\
        $m_\mathrm{\star,max}$ & 1.5 [0.11] & \multicolumn{2}{c}{$\,\,\,<4.8$} & \multicolumn{2}{c}{$\,\,\,<7.0$} \\
        $m_\mathrm{sub,max}$  & 6.1 [0.63] & 3.3 &  8.4 & 0.4 &  9.7 \\
        $m_\mathrm{gas,max}$  & 8.2 [1.07] & 6.2 &  9.7 & 2.1 & 10.7 \\
        [0.5ex]
    \hline
    \end{tabular}
\end{table}

We summarize the median and 50\percent\ and 90\percent\ ranges for all characteristic times in \Cref{t:times}.
For instance, the time elapsed between being first labelled a satellite of its current host, $\tinfall$, and definitively so, $\tacc$, is roughly 1 Gyr. That is, some galaxies entered a halo in the recent past (typically within the last Gyr), have temporarily left it, and have not had the time to re-enter their host (once they reach apocentre) by $z=0$, so-called backsplash galaxies \citep{Ludlow2008, Diemer2021, Fong2021}. Backsplash galaxies are labelled as centrals by \hbt\ at $z=0$ and are therefore excluded from our sample of satellites. We make no attempt to include them. 

We take a closer look at these characteristic times in \Cref{f:times}, which shows joint histograms in the bottom-left corner plot and the logarithm of the mean subhalo-to-stellar mass ratio at $z=0$ in the upper-right corner, as a function of pairs of lookback times. A few interesting features can be noted. For instance, most panels in the bottom-left triangle show that most galaxies occupy the $y\approx x$ zone---that is, most of these events happen within a few hundred Myr of other events labelled here: a present-day satellite typically spent little or no time as satellites of other host ($\tacc\approx\tinfall\approx\tcent\approx\tsat$). The time of maximum subhalo mass is also highly correlated with these events, most notably with the last time a satellite was labelled a central. This is expected since, no matter the details, after this time a galaxy is or has been under the influence of a larger host and can therefore be subject to mass loss processes.

There are three notable exceptions to this strong correspondence between events. The first is the time of maximum gas mass, which is $\tmgas^\mathrm{lookback}\sim6-10\,\mathrm{Gya}$ mostly independent of other events. Like $\tmsub$, $\tmgas$ is most (though still quite poorly) correlated with $\tcent$. The most important relation between maximum gas mass and other characteristic times is that $\tmgas$ is the first event to take place for the vast majority of galaxies.
Given that stellar mass continues to grow well past this time,  the reason for the early $\tmgas$ is not so much gas stripping but the termination of the gas supply. While some level of quenching is observed in low-mass host \citep[e.g.][]{Brown2017,Zinger2018}, this is consistent with the finding in some simulations (including \eagle) that it requires a massive host to fully shut down star formation \citep[e.g.][]{Pallero2022,Hough2023}.

The second exception is the time of maximum stellar mass, which is by and large the last event to take place. When this happens can be divided into two populations: most galaxies attained their maximum stellar mass almost 3 Gyr after their maximum subhalo mass (although this difference increases at later times); and a second population---15\percent\ of satellites---have maximum amounts of stellar mass today, i.e. $\tmstar^\mathrm{lookback}=0.0$ Gya. This population is naturally interpreted as consisting of satellites that have not yet reached their maximum stellar mass, and as shown in \Cref{f:times} can have fallen into their present host as much as 6 Gya. The median infall time of these satellites is 2.3 Gya, compared to 6.2 Gya for the rest of the population; similarly, they were last classified as centrals 2.5 Gya compared to 7.0 Gya for the rest of the population. These `maximum stellar mass' satellites are four times as massive (in total subhalo mass) and reside in clusters 44\percent\ as massive (median values) as the rest of the population. This results from a combination of these satellites being preferentially more massive while also having spent less time under the influence of a massive host. As seen in the diagonals of \Cref{f:times}, the time of maximum stellar mass displays the largest difference between satellites in low- and high-mass clusters today: the fraction of satellites with maximum stellar masses at $z=0$ is 10\percent\ in high-mass ($\mtwo>10^{14}\,\Msun$) clusters and 42\percent\ in low-mass ($\mtwo<5\times10^{13}\,\Msun$) clusters, and the median time of maximum stellar mass occurs 4.5 Gyr earlier for satellites currently residing in massive clusters compared to those in low-mass clusters. These trends are qualitatively consistent with trends for quenched fractions found by e.g. \cite{Park2023}.

The third exception concerns $\tsat$: 906 satellites---a full 32\percent\ of the present-day satellite population---became a satelllite more than 12 Gya (i.e. at most 1.3 Gyr after their formation in the simulation). Unlike the case for the satellites currently at their maximum stellar mass time, this `early satellite' population does not appear any different from the rest of the satellite population: even though their $\tsat$ happened 5 Gyr earlier than the general population, no other characteristic time follows a significantly different distribution from the rest of the population, with the mean characteristic times being different by more than 0.1 Gyr, including $\tcent$. Similarly, both their total and stellar masses are statistically identical to those galaxies that became satellites for the first time later on, both at the present day and at all milestones; they reside in hosts with the same median mass today; and occur in the same proportion in low- and high-mass clusters. All these early satellites were born as centrals, and while for many of them it appears the assignment as a satellite may be an algorithmic glitch---after only one or two snapshots they return to being centrals for a long time---this is not true for all of them. We therefore choose to leave $\tsat$ as is for these galaxies. However, its peculiar distribution must be kept in mind in any discussions involving $\tsat$.

There is also a notable spike in $\tacc$ roughly 1.1 Gya (partially smoothed out by the kernel in \Cref{f:times}), where 12\percent\ of the entire satellite population definitively became members of their current host within a single snapshot. This spike is dominated by two major merger events: one where 115 subhaloes definitively became satellites of \texttt{HostHaloId 1}, currently hosting 239 satellites and with $\mtwo=2.1\times10^{14}\,\Msun$, and another one where 88 subhaloes definitively became satellites of \texttt{HostHaloId 5}, currently hosting 103 subhaloes and with $\mtwo=1.9\times10^{14}\,\Msun$.\footnote{All of these numbers refer to satellites with $\mstar>10^9\,\Msun$ only.} In fact, the accretion (or final infall) times in massive clusters follow a much more punctuated distribution than the overall population, meaning major mergers are more relevant overall, although this is influenced by small number statistics; as already mentioned, there are only 12 clusters with $\mtwo>10^{14}\,\Msun$.

We also note that $\tinfall$ shows a higher correlation with both the times of maximum subhalo and stellar mass than $\tacc$, which is why---coupled with the very high correlation of $r=0.86$ between $\tinfall$ and $\tacc$---we choose to focus the later discussion on $\tinfall$ and ignore $\tacc$ from now on. In other words, once a galaxy has been identified as a satellite of its current host it starts to become influenced by it; whether at some later time \hbt\ deems it to have left its host for some time \citep[which, as discussed by][is probably a true feature of membership to a cluster by most definitions]{Diemer2021} is of secondary importance regarding the galaxy's mass content.

The upper-right corner of \Cref{f:times} shows the mean present-day stellar-to-subhalo mass ratio depending on pairs of characteristic times.\footnote{This should not be confused with the mean characteristic times of satellites with a given $\msub/\mstar$.} The largest gradients in mean present-day stellar-to-subhalo mass ratio are seen in panels involving time of maximum subhalo mass, which is closely related to $\tcent$ (but the former can be expected to depend less on subhalo finder details). We might expect a similar dependence on $\tsat$ but that is not as clear due to the prevalence of early-$\tsat$ satellites already discussed. It is also apparent that recent infallers have higher mean subhalo-to-stellar mass ratios, something we discuss at length next.

In \Cref{f:times}, we also show the marginalized distributions of satellites in the most massive ($\mtwo>10^{14}\,\Msun$) and the least massive ($10^{13}\,\Msun<\mtwo<3\times10^{13}\,\Msun$) clusters in our sample. There are 12 and 142 such clusters and 1,494 and 1,059 satellites in each respective sample. As already discussed, only the time of maximum stellar mass shows a significant dependence on host cluster mass.

\subsection{Pre-processing stars and dark matter}
\label{s:preprocessing}

\begin{figure*}
    \centering
    \includegraphics[width=0.48\linewidth]{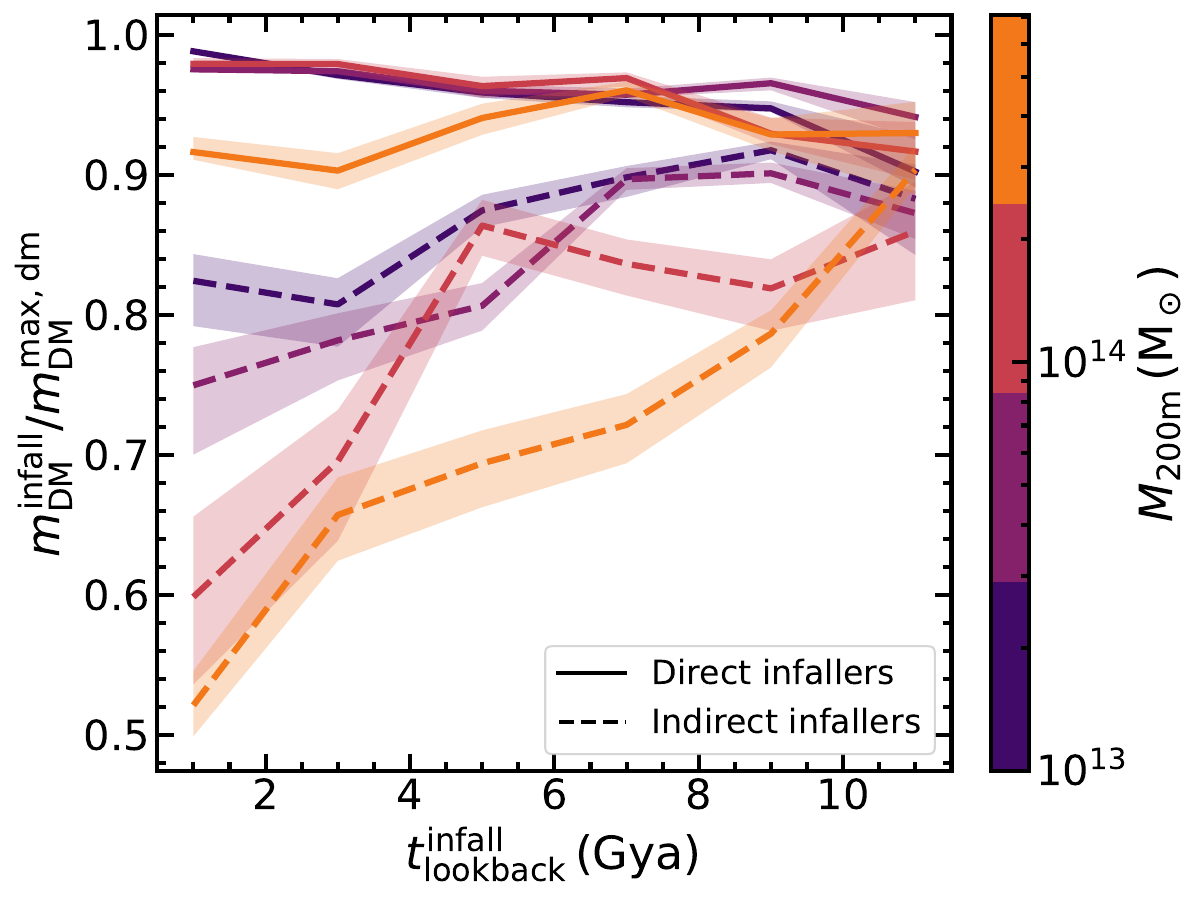}
    \includegraphics[width=0.48\linewidth]{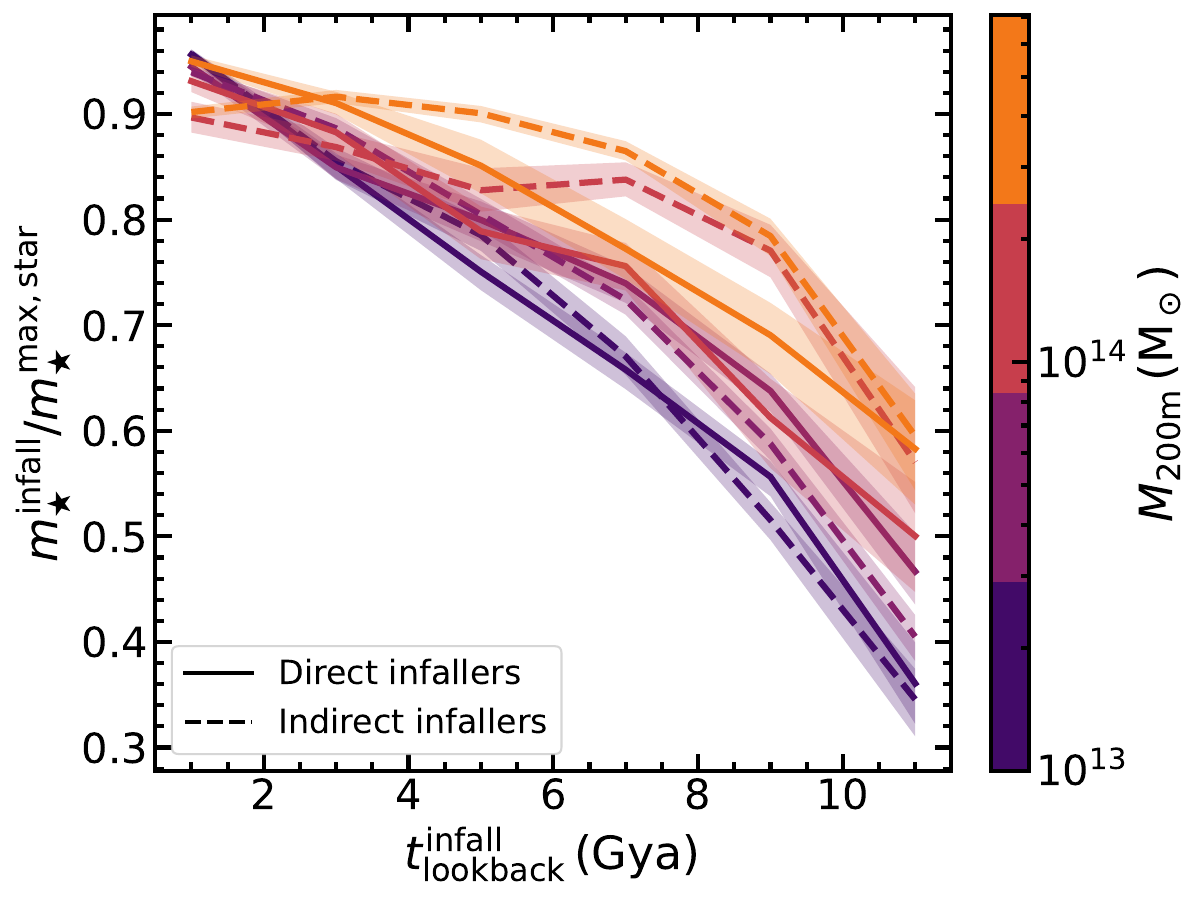}
    \caption{Pre-processing stars and dark matter. Ratio of infall to maximum dark matter (\leftpanel) and stellar (\rightpanel) masses for satellite galaxies, as a function of infall time. Colours indicate bins in present-day host halo mass, $\mtwo$. The solid and dashed lines show means of galaxies that fell in as centrals (direct infallers) and as satellites of an infalling group (indirect infallers), respectively. The  vertical scales are different.
    }
    \label{f:preprocessing}
\end{figure*}

\begin{figure*}
    \centering
    \includegraphics[width=0.48\linewidth]{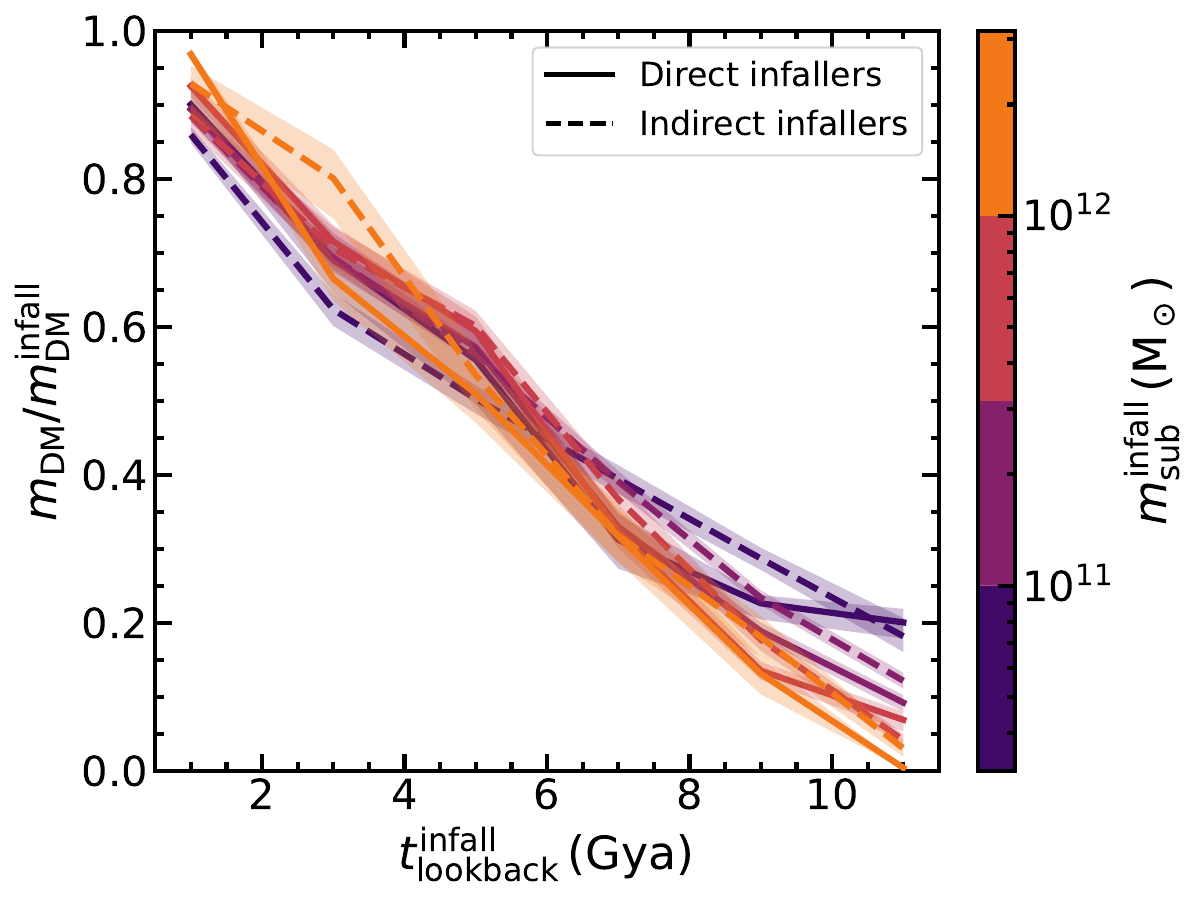}
    \includegraphics[width=0.48\linewidth]{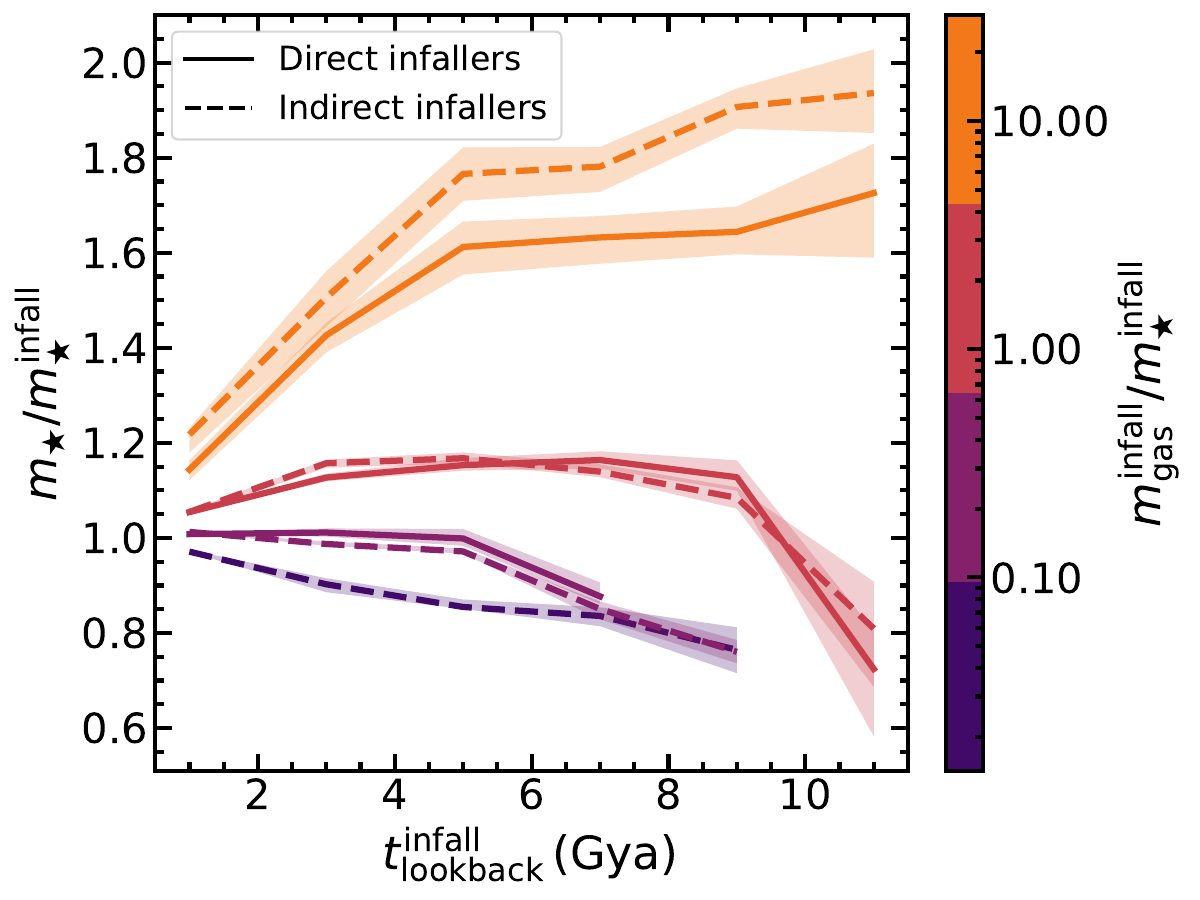}
    \caption{Post-processing stars and dark matter. \Leftpanel: Ratio of present-day to infall dark matter mass as a function of time since infall, colour-coded by total subhalo mass at infall. \Rightpanel: Ratio of present-day to infall stellar mass as a function of time since infall, colour-coded by infall gas-to-stellar mass ratio. In each case the quantity shown in the colour bar is the one that maximizes differences in behaviour. Line styles are as in \Cref{f:preprocessing}.
    }
    \label{f:postprocessing}
\end{figure*}

The concept of pre-processing is usually applied to the quenching of star formation prior to infall onto the current host.
It is generally accepted that pre-processing plays a central role in shaping cluster galaxy populations, with evidence found both in observations \citep{Haines2015,Olave-Rojas2018} and simulations \citep{Bahe2019,Pallero2022}. It is also known that the domain of influence of a dark matter halo is much larger than the virial radius, with its growth region bounded by the depletion radius~\citep{Fong2021,LiHan21,Fong22, Gao2023,Zhou2023}.
Here we discuss whether and to what extent galaxies gain or lose mass separately in the form of stars or dark matter, independent of the mechanisms, before falling into their current host.

As with star formation quenching, the mass growth of a galaxy can be significantly slowed down or even stopped well before it entered its current host---in some cases, any host at all. As summarized in \Cref{t:times}, maximum subhalo mass occured on average 1 Gyr before infall; as shown in \Cref{f:times} in some cases maximum subhalo mass can occur up to 8 Gyr before infall, and in a few rare cases it even happens before a galaxy is ever labelled a satellite of any host.  
The left panel of \Cref{f:preprocessing} shows this dark matter pre-processing more directly: on average, satellites in massive ($\mtwo>10^{14}\,\Msun$) clusters fell into their current host with 70--80\percent\ their maximum dark matter mass, and as low as 60\percent\ at $z=0$. Even satellites currently residing in low-mass clusters ($10^{13}<\mtwo/\Msun<10^{14}$) fell in to the present host with roughly 90\percent\ their maximum dark matter mass.\footnote{These overall numbers can be extracted visually from \Cref{f:preprocessing} by combining the solid and dashed lines accounting for uncertainties, since both sets are mutually exclusive. The curves have been omitted from the figure to avoid overcrowding.} 
In fact, the level of pre-processing depends strongly on whether a present-day satellite fell in as a central or as a satellite in a smaller host. We use the terms `direct infall' and `indirect infall' to refer to present-day satellites that fell into their current host as central or satellite galaxies, respectively. In the latter case, the subhalo became a sub-subhalo,\footnote{Sub-subhaloes are identified in \hbt\ as having \texttt{Depth} $>1$.} at least temporarily. At any given time, between 30 and 40\percent\ of infallers in \eagle\ do so as satellites of an infalling group, i.e. through indirect infall. Galaxies that fell in directly (i.e. as centrals) did so with almost their maximum dark matter masses. The exception is directs infallers falling into the most massive clusters, which fell in having lost almost 10\percent\ of their dark matter mass on average while being centrals. We may expect a small level of dark matter loss even for central galaxies, because tidal stripping can act well before a central halo reaches the virial radius of its neighbour \citep{Behroozi2014}, or due e.g.\ to relaxation following a major merger or tidal losses from close encounters \citep{Lee2018}. Indirect infallers, in contrast, lost as much as 40--50\percent\ of their dark matter to pre-processing, on average. These trends are also consistent with results in \nbody\ simulations \citep[e.g.][]{Joshi2017}. As expected, the amount of dark matter pre-processing increases with cosmic time for galaxies falling into their present hosts indirectly as satellites of another host. The ratio of dark matter at infall to maximum dark matter is constant with time for direct infallers in the most massive clusters, but it increases with cosmic time for clusters with $\mtwo\lesssim10^{14}\,\Msun$, as neighbouring gravitational potentials get stronger.\footnote{We note that some galaxies do reach maximum subhalo mass \emph{after} infall. However, this is not the dominant behaviour for direct nor indirect infallers at any given time. On average, therefore, it is appropriate to interpret the lower infall subhalo mass as pre-processing rather than as late growth.}

Stellar mass shows the opposite trend. As shown in \Cref{f:times}, it is quite rare for a galaxy to reach maximum stellar mass before infall; only 12\percent\ of present-day satellites had $\zmstar>\zinfall$. In fact, for galaxies with $\tinfall^\mathrm{lookback}\gtrsim3$ Gya maximum stellar mass typically occurs 2 Gyr after infall to the present host. More recent infallers cover a wide range of $\tmstar$, with the latter tending to happen earlier, simply because $\tinfall$ happens late. As with dark matter mass, both the maximum stellar mass and the trend with infall time depend on both (present-day) cluster mass and on whether the galaxy fell in directly or indirectly, although the latter dependence is seen only for massive ($\mtwo>10^{14}\,\Msun)$ clusters. In particular, galaxies that entered their current host, say, 8 Gya, did so on average with 60--75\percent\ their maximum stellar mass depending on cluster mass. (We have verified that the dependence on cluster mass is more significant than that on subhalo mass.) As already discussed, maximum stellar mass by and large happens after infall; this means satellites in lower-mass clusters grew almost twice as much stars after infall than their counterparts in higher-mass clusters. However, galaxies that fell into $\mtwo>10^{14}\,\Msun$ clusters indirectly gained no more (on average) than 20\percent\ additional mass afterwards. The fact that this maximum gain has remained constant over the last 7--8 Gyr (i.e. since $z\simeq1$) for indirect infallers only is another way of seeing star formation quenching by pre-processing. In contrast, the post-infall stellar mass gain continues to increase even to the present day for indirect infallers into low-mass clusters.

\subsection{Stars and dark matter after infall}
\label{s:postprocessing}

\Cref{f:postprocessing} shows what happens after \Cref{f:preprocessing}: upon infall, a galaxy steadily loses dark matter, taking an average 5 Gyr for a galaxy to lose half the dark matter it had at infall. The left panel of \Cref{f:postprocessing} shows that the dark matter loss rate since infall, at fixed infall time, depends slightly on infall mass. Among recent infallers ($\tinfall<2$ Gya), less massive subhaloes have lost a fraction of dark matter approximately twice as high as the more massive subhaloes. However for the very early infallers ($\tinfall\sim10$ Gya) the more massive subhaloes have lost higher dark matter fractions. Although the trend over time does not depend on cluster mass, satellites in lower-mass clusters always retain roughly ten percentage points more of their infall dark matter (not shown). For instance, satellites that fell 10 Gya to a cluster with present-day mass $\mtwo\sim2\times10^{13}\,\Msun$ retain 15--20\percent\ of their infall dark matter, while sateliltes that fell 10 Gya to a cluster with present-day $\mtwo\sim2\times10^{14}\,\Msun$ retain 5-10\percent. Similarly, satellites that fell in to their present host 2 Gya retain, on average, 85\percent\ (75\percent) their infall dark matter mass if they reside in a $\mtwo\sim2\times10^{13}\,\Msun$ ($\mtwo\sim2\times10^{14}\,\Msun$) cluster. Whether a subhalo fell in directly or indirectly is only apparent in terms of dark matter loss within the first 2--3 Gyr after infall; the dark matter fraction lost since infall does not depend on the status at infall for earlier infallers. This timescale can be interpreted as the timescale it takes the new host to take over the small-scale gravitational influence over the satellite galaxy, and may in fact be associated with the dismembering of the infalling host \citep{Haggar2023}.

The right panel of \Cref{f:postprocessing}, on the other hand, shows that stellar mass continued to grow significantly after infall. This could already be seen from \Cref{f:times}; \Cref{f:postprocessing} shows that this is true specifically for satellites that fell in to their present host with $\mgas^\mathrm{infall}\gtrsim\mstar^\mathrm{infall}$, independent of when they fell in. It is quite interesting that very gas-rich galaxies, with $\mgas^\mathrm{infall}\gg\mstar^\mathrm{infall}$, gained a larger stellar mass fraction if they fell in indirectly than if they fell in directly, independently of the time of infall.
For galaxies to have less stellar mass today than they had at infall requires either a long time since infall, $\tinfall\gtrsim10$ Gya, or a low availability of gas at infall, $\mgas^\mathrm{infall}\ll\mstar^\mathrm{infall}$.

\begin{figure*}
    \centerline{\includegraphics[width=\linewidth]{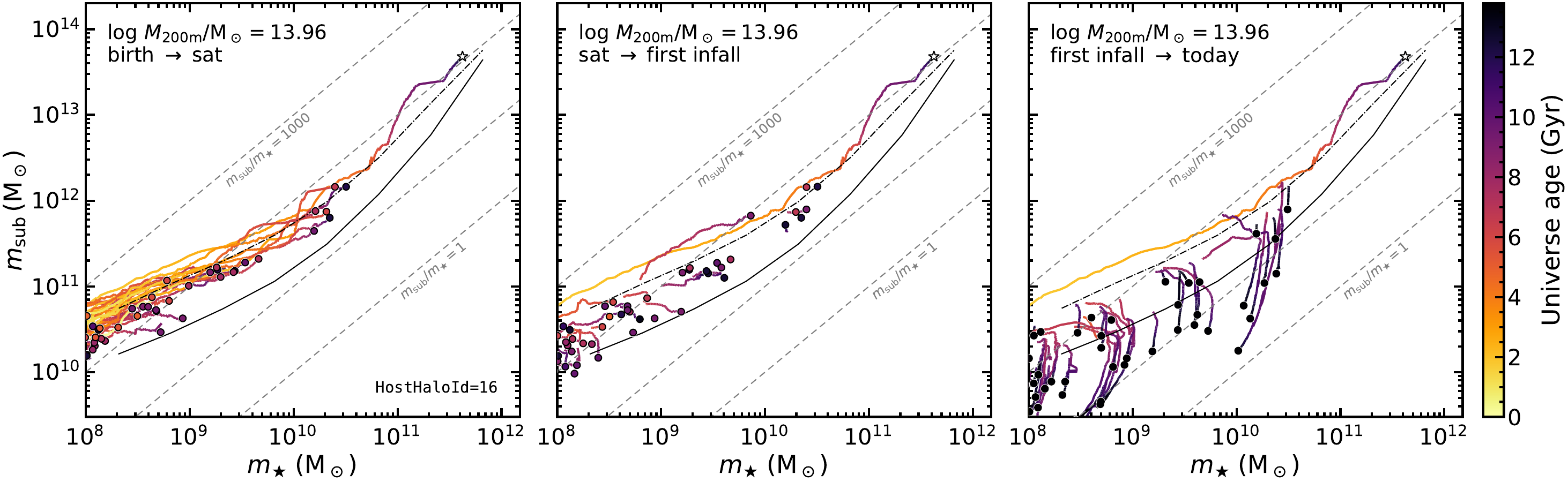}}
    \centerline{\includegraphics[width=\linewidth]{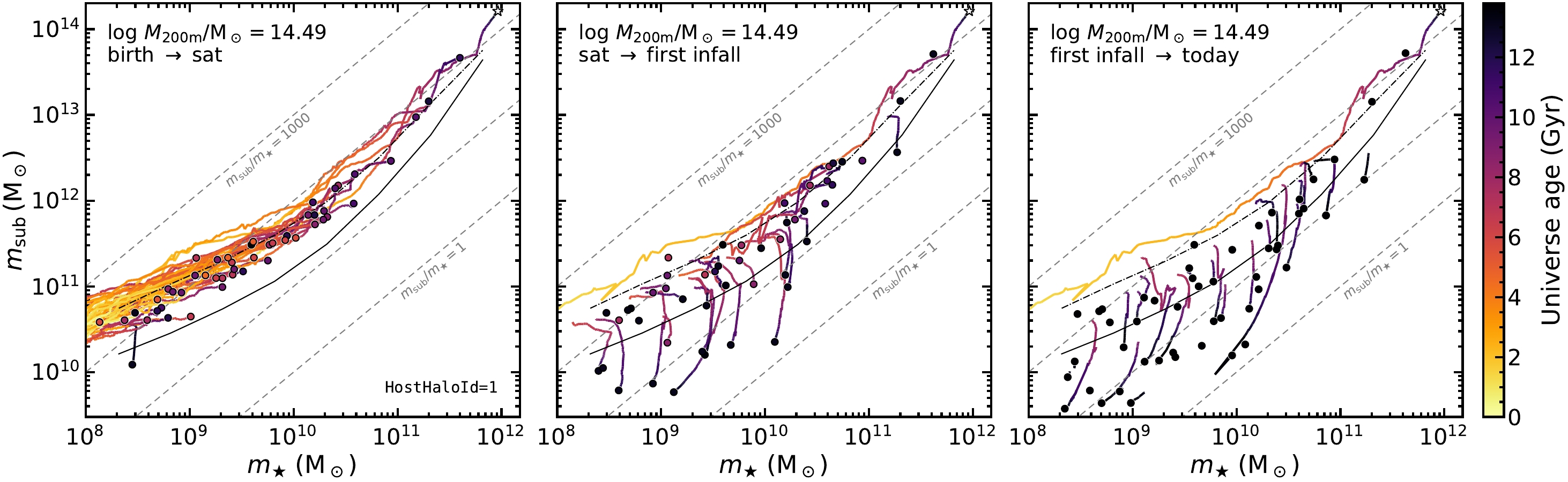}}
    \caption{Evolution in the subhalo mass-stellar mass plane within two randomly selected clusters in EAGLE, one in the low-mass end and the other in the high-mass end of our sample (\texttt{HostHaloId 16} and \texttt{1} at $z=0$ in the \hbt\ catalogue, respectively). 
    From \leftpanel\ to \rightpanel, we show the motion in the $\msub-\mstar$ plane of 38 and 50 subhaloes within each cluster, respectively, selected randomly to span the full stellar mass range explored in this work but including the six most massive subhaloes (including the central subhalo, whose track ends on a white star), and only selecting satellites with $\tsat<12$ Gya (see \Cref{s:milestones}). Subhaloes with present-day stellar masses $\mstar<10^9\,\Msun$ are shown here for illustration, but are not part of the sample analysed in this work. All three panels show the same subhaloes per host. As labelled in the top left, each panel shows the time span between the birth of a subhalo and the first time they were labelled a satellite of any host (\leftpanel), since then until they fell into their present host for the first time (\midpanel), and from first infall until the present day (\rightpanel), except for the track of the central subhalo which is reproduced from birth to the present day in all panels. Each subhalo's history is colour-coded by age of the Universe. The thin solid and dash-dotted black lines show the present-day satellite and central SHSMR, respectively, for reference. The grey dashed diagonal lines show constant subhalo-to-stellar mass ratios, increasing by factors of ten from bottom to top, as labelled.
    }
    \label{f:hsmr_buildup}
\end{figure*}

\begin{figure*}[h!]
    \centering
    \includegraphics[width=\linewidth]{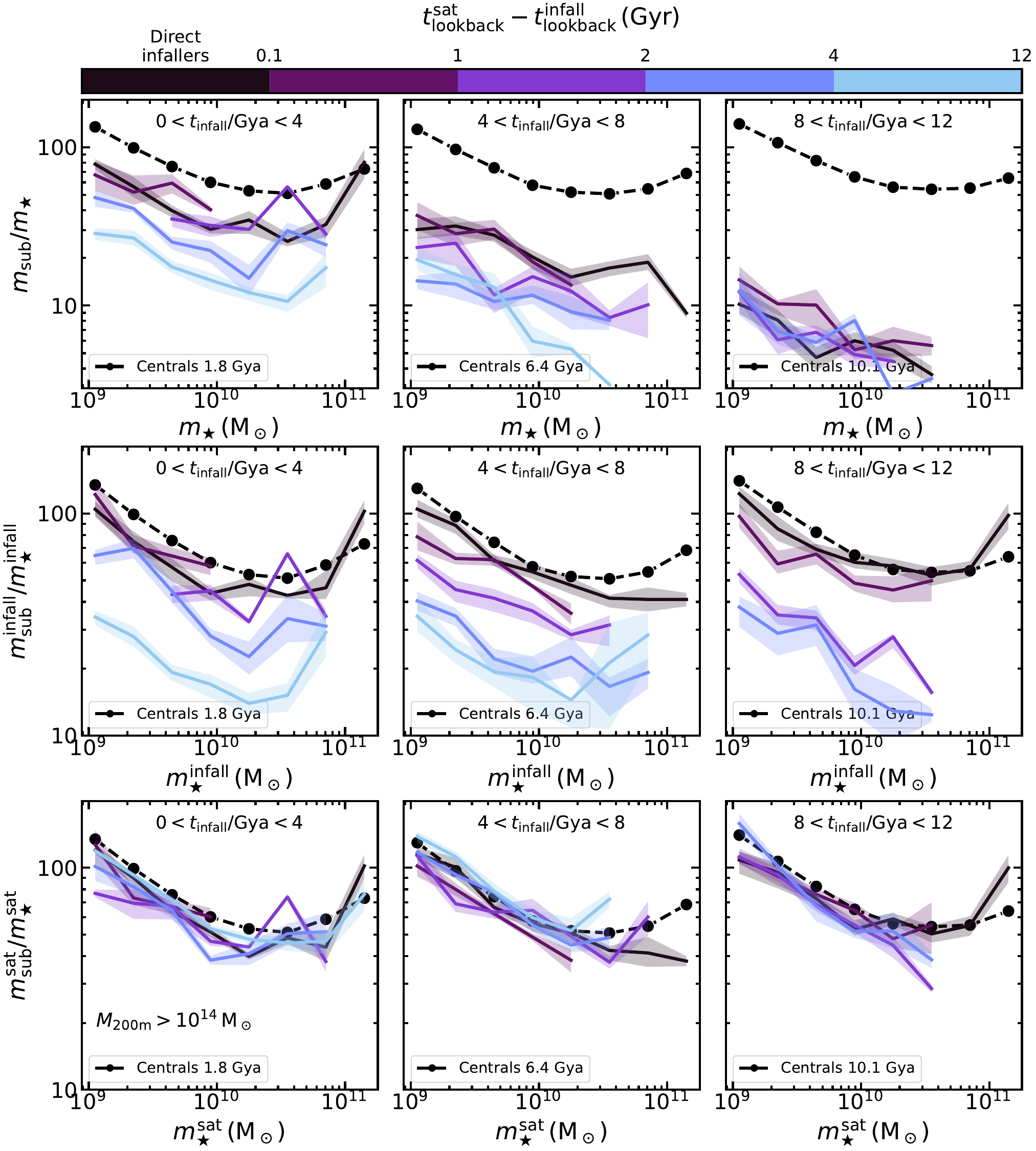}
    \caption{Pre- and post-processing of the satellite SHSMR. From \toppanel\ to \bottompanel\ are shown the SHSMR at $z=0$, at the time of infall to the present host and at the time each galaxy was labelled a satellite for the first time. From \leftpanel\ to \rightpanel\ are shown increasing times since infall, as labelled. The black solid lines show the SHSMR for galaxies that fell in to their present hosts as centrals, while the coloured lines show galaxies that fell in as satellites of another host, colour-coded by time elapsed between first being labelled a satellite and falling into the present host, for satellites residing in $\mtwo>10^{14}\,\Msun$ only. The black dashed lines with circles show the SHSMR of central galaxies at the median infall time of each sample. The shaded regions show normal uncertainties on the mean. The  vertical scale is different in each row.}
    \label{f:hsmr_evolved}
\end{figure*}

\section{Evolving the satellite SHSMR}
\label{s:hsmr_evol}

Having discussed the history of satellites and their mass content at length, we now turn to the build-up of the SHSMR and its relation to the central SHSMR today.

\subsection{Evolution of the mean relation}
\label{s:hsmr_evol_mean}

\Cref{f:hsmr_buildup} illustrates the evolution of the mass content of subhaloes in detail for two randomly selected clusters, one on the low-mass end and another on the high-mass end of our sample, in terms of the milestones described in \Cref{s:milestones}. The most striking difference between satellites in either cluster is when subhaloes suffer most of the mass loss: in the more massive cluster much of the mass loss happens prior to infall when satellites resided in another host (i.e. in the middle panel), while in the lower-mass cluster most of the mass loss suffered by subhaloes happens within their current host (i.e. in the rightmost panel). 
Some subhaloes in the massive host lost a significant fraction of their total mass before being satellites of any host at all, and many spent a significant amount of time as satellites before falling into their current host: they fell into the massive cluster indirectly as satellites. On the contrary, most galaxies in the low-mass cluster do not have a long trail in the middle panel, i.e. $\tsat\approx\tinfall$. Many galaxies in the low-mass cluster also kept growing their stellar mass for a significant time after infall, while this is rarely the case in the massive cluster.

Once mass loss begins it happens quickly; some of the latest-falling galaxies already lost an order of magnitude in total mass within the last 2 Gyr. This may be due to a number of effects. For one, the host cluster is continuously growing over time, which means the potential encountered by late infallers is much deeper---the gravitational effects significantly stronger---than that encountered by early infallers (halo growth can be traced approximately by the growth of the central subhalo depicted in \Cref{f:hsmr_buildup}). However, stellar mass usually continues to grow well after total mass starts to decrease, in particular for early infallers. As discussed in \Cref{s:milestones}, galaxies that fell early on may have still had a significant gas reservoir, while late infallers had already lost their gas supply and were not able to grow their stellar mass as much.
The trends of \Cref{f:preprocessing} are manifested in the fact that most tracks are already ``moving'' down in the bottom middle panel---decreasing $\msub$ before infall to the massive cluster---and that most tracks in the right-most panels have some rightward movement (i.e. increasing $\mstar$ after infall), although less so in the massive cluster (bottom panel). The same behaviour has been described in the IllustrisTNG simulation by \cite{Engler2021}.

\begin{figure*}[h!]
    \centering
    \includegraphics[width=\linewidth]{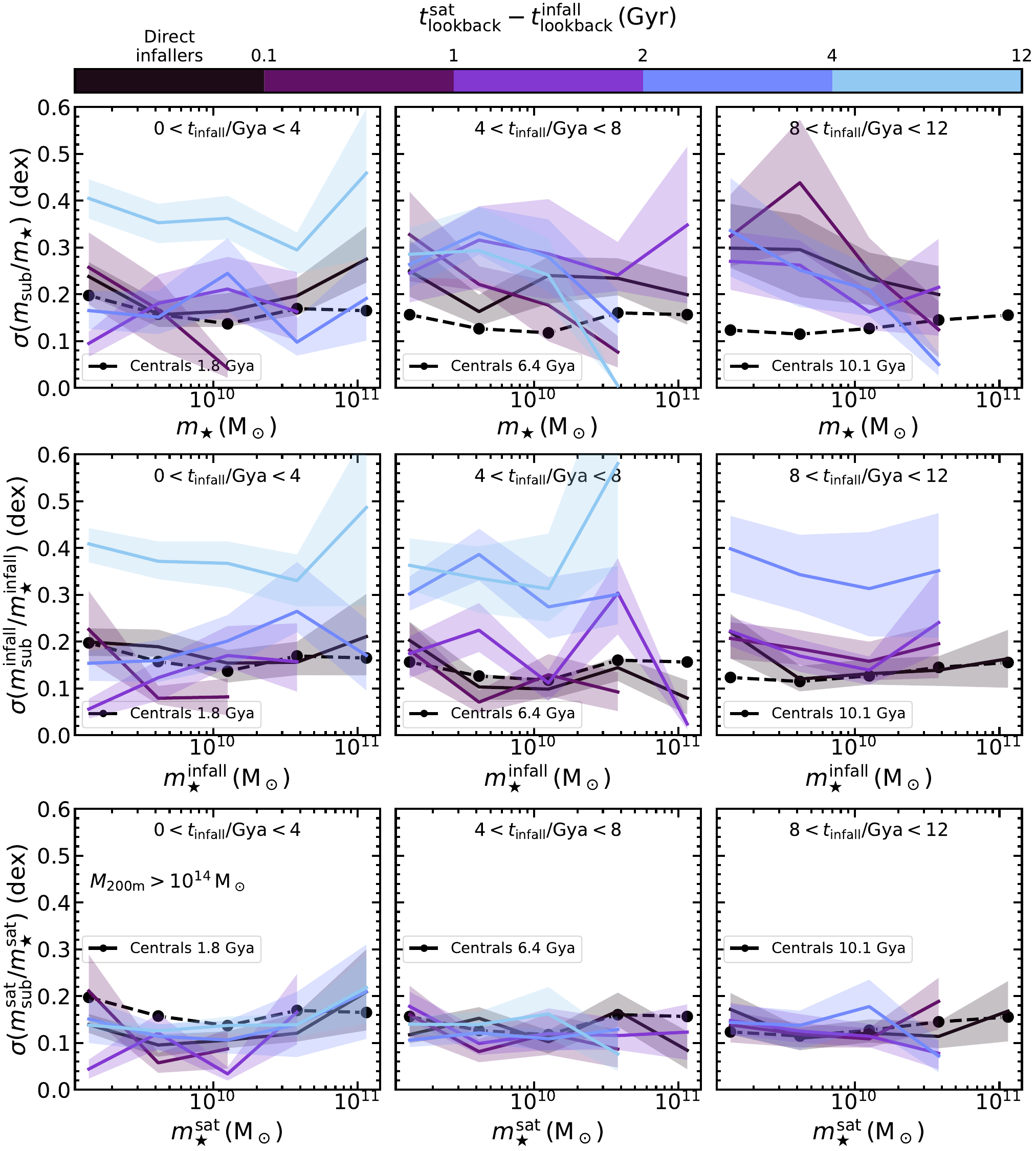}
    \caption{Impact of pre- and post-processing on the satellite SHSMR scatter over time. Samples, binning, and styles are as in \Cref{f:hsmr_evolved}.}
    \label{f:scatter_evolved}
\end{figure*}

The impact of the combination above in the SHSMR over time is shown in \Cref{f:hsmr_evolved}. Anticipating the discussion, we note two differences with respect to the rest of the paper, which also apply to \Cref{f:scatter_evolved}. First, we only show the SHSMR of satellites in massive clusters, $\mtwo>10^{14}\,\Msun$, to avoid cluttering. For lower-mass hosts we see a reduction in the impact of pre-processing, which for $\mtwo<5\times10^{13}\,\Msun$ hosts is approximately half that discussed below, as shown in \Cref{f:preprocessing_smallhost}. Second, we include all satellites regardless of their present mass in each panel, i.e. we include satellites that have gone below $\mstar=10^9\,\Msun$ at $z=0$ but have large enough masses at infall or when first becoming a satellite, as appropriate for each panel. This latter change makes the discussion less sensitive to artificial disruption \citep[e.g.][]{Han2016,Bahe2019}.\footnote{We also exclude galaxies with $\tsat>12$ Gya, which reduces the statistical power of these results but, as discussed in \Cref{s:milestones}, has no impact on our conclusions.}
The top row shows that the present-day SHSMR depends on pre-processing time but only for galaxies infalling in the last 8 Gyr. It appears as though it takes at least 2 Gyr for pre-processing to become apparent in the SHSMR at late times ($\tinfall<4$ Gya). At any rate, the  dependence of the SHSMR on time since infall is at least comparable to the dependence on pre-processing time. This is consistent, for instance, with recent results on galaxy quenching in IllustrisTNG \citep{Park2023} and as inferred in observations of clusters at $z\sim1$ \citep{Werner2022}. The shorter extent of the dark purple line in the top-left panel means, roughly speaking, that there are no massive satellites ($\mstar>10^{10}\,\Msun$) which became satellites for the first time in the last 4 Gyr and then fell into their current host within 1 Gyr.

The middle row shows the pre-processing of the SHSMR directly as the SHSMR at the time of infall, binned in an identical manner as discussed above. Galaxies with longer pre-processing times fall into their current hosts with significantly lower SHSMRs, and the dependence with time since infall is mostly removed, as expected. Here it is clear that 1 Gyr is enough pre-processing time to significantly shift the SHSMR of galaxies falling into massive clusters (at least until roughly 4 Gya).

Finally, the bottom panels of \Cref{f:hsmr_evolved} show the SHSMR at the time present-day satellites first became satellites of any host, i.e. $\tsat$. The left panel shows, surprisingly, that satellite galaxies have a lower SHSMR than central galaxies even when they first become satellites. This shows that will-be satellite galaxies are influenced by the larger-scale environment differently than galaxies that will remain centrals up to the present day. There is a hint that this effect may be stronger for galaxies with shorter pre-processing times; however, this is not statistically significant. The lower SHSMR of will-be satellites is even apparent at the time of maximum subhalo mass (not shown) instead of $\tsat$, showing this is not an artefact of the labelling of a galaxy as satellite by \hbt.
In words, dark matter pre-processing is more dramatic than suggested by the discussion in \Cref{s:preprocessing}: before removing mass it halts growth. Galaxies that will become satellites in the future have already began being pre-processed by the time they reach maximum mass. This has important implications for the estimation of mass loss observationally, which we discuss in \Cref{s:massloss_obs}.

As mentioned, \Cref{f:preprocessing_smallhost} shows the pre- and post-processing of the SHSMR for satellites currently residing in low-mass clusters. The levels of pre-processing are smaller, as may be expected, but the trends are otherwise equivalent. It is noteworthy that the SHSMR at $\tsat$ is also lower than that of centrals at the same times.

\subsection{Evolution of scatter in the SHSMR}
\label{s:hsmr_evol_std}

We now turn to the evolution of the scatter in the satellite SHSMR, which we explore in \Cref{f:scatter_evolved}. The first interesting fact from these figures is that when controlling for satellite history the scatter in the satellite SHSMR is significantly reduced. This is already apparent in \Cref{f:hsmr_evolved}, in that  $\tinfall$ clearly separates occupation in the $\msub-\mstar$ plane. The scatter in the present-day SHSMR decreases slightly for more recent infallers, and except for those galaxies with the longest pre-processing times it is independent of pre-processing time itself. This suggests, as does the discussion below, that it is the time spent as a satellite, $\tsat$, the time that dominates the present-day SHSMR scatter. Whether the time spent as a satellite is spent within a less massive or more massive host is of secondary importance in these terms.

As the bottom panels show, when becoming satellites the SHSMR shows as little scatter as that of centrals. The middle panels, in turn, isolate the effect of pre-processing, which increases the scatter by up to 0.2 dex, with a timescale that depends on infall time (or, equivalently for this purpose, on $\tsat$). That is, early infallers suffer pre-processing more quickly than late infallers, which seem to require several Gyr for the scatter to increase. This increase in scatter is, in the end, equally large as for recent infallers. As anticipated this suggests that $\tsat$ is more important than pre-processing time in establishing both the mean and scatter of the SHSMR at infall.

As in the preceding section, we note that \Cref{f:scatter_evolved} only shows satellites currently residing in $\mtwo>10^{14}\,\Msun$. \Cref{f:preprocessing_smallhost_scatter} shows the equivalent for satellites in $\mtwo<5\times10^{13}\,\Msun$ clusters. The only difference may be that the scatter in the SHSMR increases more quickly due to pre-processing for satellites currently residing in low-mass clusters, but the trends are otherwise consistent.

\subsection{The evolution of segregation}
\label{s:segregation_evol}

\begin{figure}
    \centering
    \includegraphics[width=\linewidth]{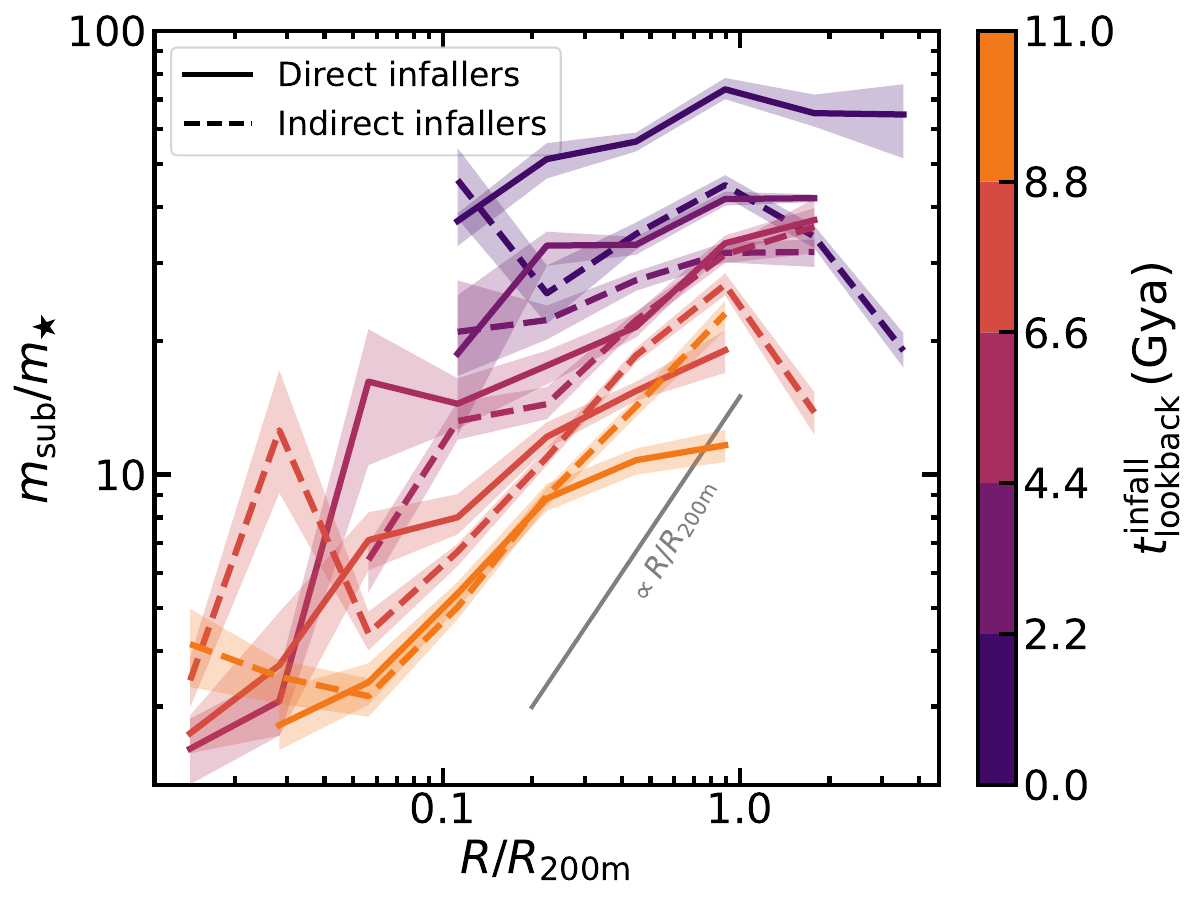}
    \caption{Mass segregation build-up. Subhalo-to-stellar mass ratio as a function of present-day three-dimensional distance to the cluster centre, normalized by $\rtwo$ and split according to time since infall, as shown in the colour bar. The solid and dashed lines show direct and indirect infallers, respectively. The grey line shows a linear relation, for reference.
    }
    \label{f:segregation_since_infall}
\end{figure}

Finally, we look at how the segregation described in \Cref{s:segregation} builds up over time. \Cref{f:segregation_since_infall} clearly shows the trend in \Cref{f:mratio_dist} evolving since infall, with galaxies both getting closer to the centre and losing preferentially more dark matter over time. Galaxies that fell most recently--within the last 2 Gyr or so---show little segregation within $\rtwo$, and cannot be found closer in than $0.1\rtwo$ as they have yet to reach the centre of the cluster.
Except for the innermost radii, 10 Gyr after infall the SHSMR depends on cluster-centric distance as $\propto(R/\rtwo)^{0.70}$. The flattening at $R<0.1\rtwo$ is due to our selection in present-day stellar mass, which means for heavily stripped subhaloes we only retain those with high $\msub/\mstar$. We confirm that selecting by infall mass instead produces relations consistent with power laws in the entire $R<\rtwo$ range, but the slopes at $R\gtrsim0.1\rtwo$ do not change significantly.
Using $\tsat$ instead of $\tinfall$ (not shown) produces trends that match each other more closely than \Cref{f:segregation_since_infall}, i.e. segregation does not depend as strongly on $\tsat$ as it does on $\tinfall$. This is expected since pre-processing happens by definition while a galaxy is in the outskirts, or even entirely outside, of its (eventual) host cluster, well before galaxies have sunk into the cluster centre. 

Recent infallers show different subhalo-to-stellar mass ratios at fixed cluster-centric distance but this is no longer the case for $\tinfall\gtrsim7$ Gya. Of such early infallers, indirect infallers have considerably lower ratios than direct infallers, due to the effect of the infalling host. Interestingly, the observation in \Cref{f:mratio_dist} that satellites have a lower $\msub/\mstar$ beyond $\rtwo$ is due to indirect infallers alone, while direct infallers have a constant $\msub/\mstar$ from $\rtwo$ outwards. The galaxies that make it beyond $\rtwo$ after infall are those with the most elliptical orbits and therefore shortest pericentres, which suffer the most mass loss \citep[e.g.][]{Bakels2021,Smith2022}. However, \cite{Haggar2023} have shown that galaxies in infalling groups are largely shielded from the host cluster if they are within the tidal radius of the infalling group, at least for some time. Combined, these effects explain why only indirect infallers show a decrease in $\msub/\mstar$ when located outside $\rtwo$ of the host cluster. 

We also note that the fact that earlier infallers tend to be located closer to the cluster centre was discussed by \cite{vandenBosch2016}, who showed that the strong segregation of $\msub/\msub^\mathrm{infall}$ of dark matter subhaloes is due to the inside-out growth of haloes, i.e. that earlier infallers joined the host halo closer to the centre than late infallers do.

\section{Quantifying mass loss in observations}
\label{s:massloss_obs}

\begin{figure}
    \centering
    \includegraphics[width=\linewidth]{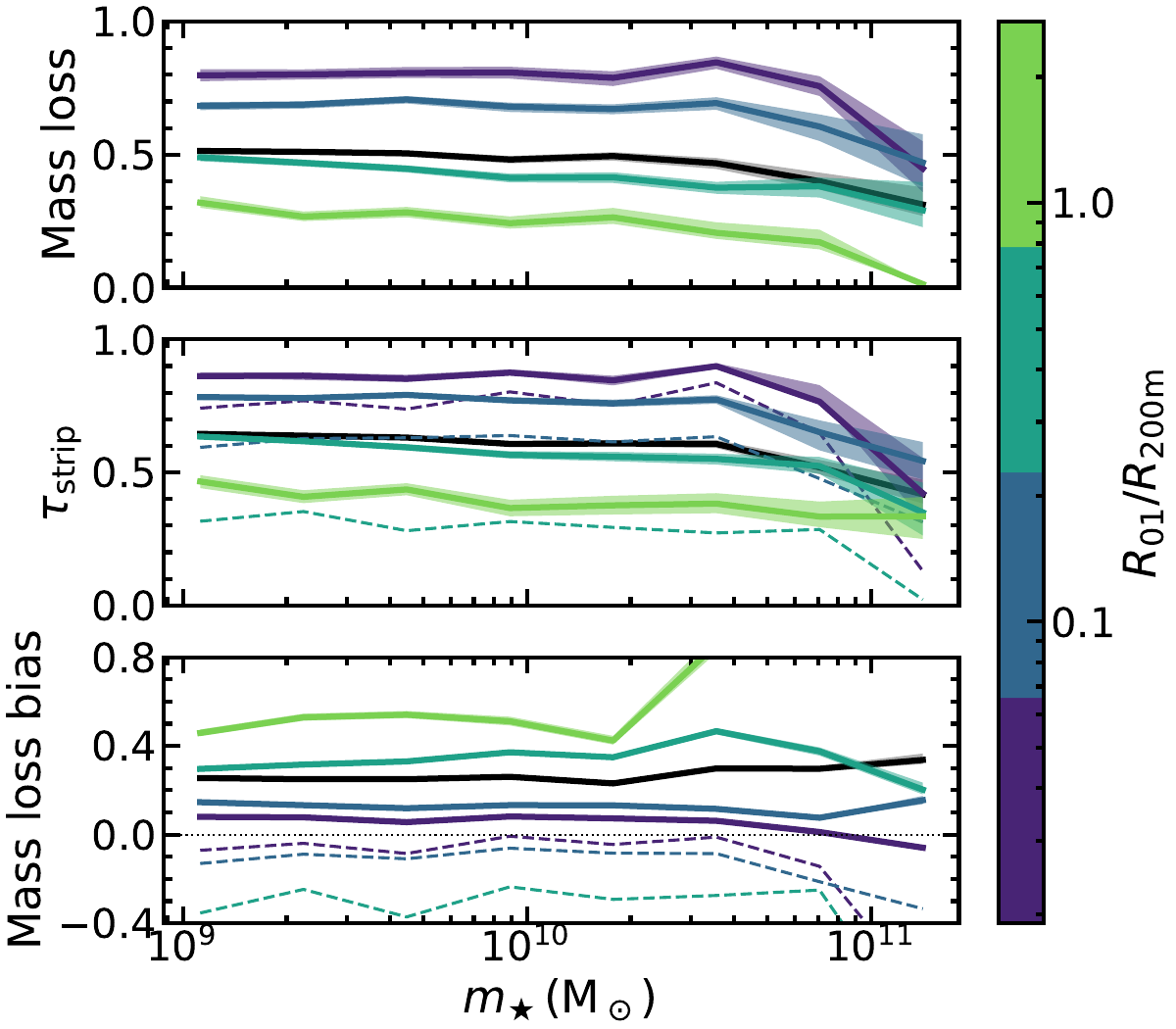}
    \caption{Bias incurred in the estimation of mass loss by assuming it can be calculated as the ratio of satellite to central subhalo mass at fixed stellar mass (\bottompanel). This bias is calculated as the ratio of the present-day to maximum subhalo mass (i.e. mass loss itself; \toppanel), and $\tau_\mathrm{strip}=1-m_\mathrm{sat}/m_\mathrm{central}$ (i.e. the observational proxy; \midpanel). All curves are colour-coded by projected cluster-centric distance normalized by $\rtwo$, and the shaded regions show corresponding normal uncertainties. The dashed lines correspond to using the outermost bin in the denominator instead; see \Cref{s:massloss_obs} for details.
    The black lines show quantities for all satellites together.
    }
    \label{f:massloss}
\end{figure}

It is common among observational studies to attempt to quantify mass loss in satellite galaxies by comparing their subhalo-to-stellar mass ratio with that of a sample of central galaxies with matching properties, usually stellar mass \citep[e.g.][]{Kumar2022}. Under the assumption (justified, as we have seen) that stellar mass is much more stable than dark matter to mass loss, differences in the SHSMR between the two samples are taken as indication of mass loss.  For instance, \cite{Wang2023} defined a stripping fraction as $\tau_\mathrm{strip}=1-\msub/m_\mathrm{central}$, i.e. assuming $m_\mathrm{central}=\msub^\mathrm{max}$ at fixed stellar mass, using the central SHSMR from \cite{Shan2017} to obtain $m_\mathrm{central}$. Alternatively, \cite{Niemiec2017} defined $\tau_\mathrm{strip}=1 - \msub^\mathrm{inner}/\msub^\mathrm{outer}$, where $\msub^\mathrm{inner}$ and $\msub^\mathrm{outer}$ correspond to the mean masses of subhaloes in the inner and outer regions of clusters, respectively.

As we have seen, stellar mass can increase quite significantly after infall. This means that the timescales for dark matter mass and stellar mass build-up in satellite galaxies are different, with maximum stellar mass attained typically 4 Gyr after maximum dark matter mass. Therefore, the points of comparison for dark matter loss and stellar mass loss are different. On the other hand, stellar mass generally contributes a small fraction of the total mass, even for heavily stripped subhaloes, so the stellar mass increase makes little difference for the overall amount of mass lost due to stripping.

With the aim of providing an anchor for observational results such as the lensing measurements discussed in \Cref{s:segregation}, here we evaluate how  the total true mass loss compares to $\tau_\mathrm{strip}$. We calculate this mass loss bias as
\begin{equation}\label{eq:massloss}
    \mathrm{bias} = \frac{\tau_\mathrm{strip}}{1-m_\mathrm{sub}/m_\mathrm{sub}^\mathrm{max}} - 1= \frac{1 - m_\mathrm{sub}/m_\mathrm{central}}{1-m_\mathrm{sub}/m_\mathrm{sub}^\mathrm{max}} - 1 
,\end{equation}
where $\msub$ and $m_\mathrm{central}$ correspond to the subhalo masses of satellite and central galaxies, respectively (including baryons and dark matter). \Cref{f:massloss} shows that this bias is approximately independent of stellar mass but strongly dependent on projected cluster-centric radius, increasing from 6\percent\ for the innermost bin shown to 51\percent\ in the outermost bin.\footnote{These values correpsond to inverse-variance weighted averages of data points with $\mstar<5\times10^{10}\,\Msun$, beyond which uncertainties dominate. In particular, note how the bias for $R_{xy}>\rtwo$ diverges at higher stellar masses, because mass loss itself approaches zero.} We also calculate $\tau_\mathrm{strip}$ similarly to \cite{Niemiec2017}, by taking the ratios between mean subhalo masses in each of the three innermost radial bins and the outermost one, covering $0.8<R_{xy}/\rtwo\lesssim2$; these lead to the dashed lines. Because such a definition essentially ignores pre-processing, it leads to an underestimation of mass loss. \cite{Niemiec2017} used for the denominator a radial bin similar to the third radial bin in \Cref{f:massloss}, namely $0.55<R/(h^{-1}\,\mathrm{Mpc})<1.0$; because EAGLE galaxies in that radial bin have already lost 40--50\percent\ of their total mass already, this would lead to a larger underestimation of mass loss.
We list the mean mass loss bias, \Cref{eq:massloss}, in \Cref{t:massloss}, when averaging each sample over the whole stellar mass range. Although the bias is significant it is well behaved in that it does not depend on stellar mass; it can therefore be treated as a correction to $\tau_\mathrm{strip}$ that depends only on the radial distribution of the satellites under consideration.

\begin{table}
    \centering
    \begin{tabular}{l | c c c c}
    \hline
    \hline
        Mean radius ($\rtwo^{-1}$) & 0.035 & 0.1 & 0.4 & 1.5 \\
        Mass loss bias & 0.06 & 0.13 & 0.32 & 0.51 \\
    \hline
    \end{tabular}
    \caption{Mass loss bias, \Cref{eq:massloss}, for each bin in projected cluster-centric distance in \Cref{f:massloss}, averaged over all stellar masses.}
    \label{t:massloss}
\end{table}

\section{Conclusions}
\label{s:conclusions}

We studied the mass content of satellite galaxies in galaxy clusters with masses $\mtwo>10^{13}\,\Msun$ using an improved subhalo catalogue produced with the \hbt\ algorithm \citep{Han2018} applied to the 100 Mpc-box \eagle\ cosmological hydrodynamical simulation, both at present and as a function of time. 
Consistent with previous work, we find satellites have a significantly reduced subhalo-to-stellar mass ratio compared to present-day central galaxies (\Cref{f:hsmr}) and roughly 3 times higher intrinsic scatter in the subhalo-to-stellar mass relation at fixed stellar mass (\Cref{f:scatter_mhost}). We find a strong dependence of the subhalo-to-stellar mass ratio on cluter-centric distance, independent of stellar mass, which is consistent with recent weak lensing measurements (\Cref{f:mratio_dist}).

We also investigated the histories of present-day cluster galaxies. 
On average, satellite galaxies have spent roughly 5 Gyr within their current host. 30-40\percent\ of surviving satellites fell in to their present host indirectly as satellites of less massive systems, and this has a significant impact on their mass content: these indirect infallers had already lost 50--70\percent\ of their dark matter by infall, while direct infallers, i.e. those which fell into their current host as central galaxies, had lost $<10$\percent. Stellar mass, in contrast, continues to grow for roughly 2 Gyr after infall, and most satellites still have higher stellar masses today than they did at infall (\Cref{f:preprocessing,f:postprocessing}).

In fact, present-day satellites have lower SHSMR than centrals even at the time of maximum subhalo mass several Gya (\Cref{f:hsmr_evolved}, lower panels), suggesting that while they were centrals they already experienced harsher environments than those galaxies which are still centrals today. Similarly, even galaxies in low-mass clusters (with $10^{13}<\mtwo/\Msun<5\times10^{13}$) show significant signs of pre-processing, with the SHSMR at infall of indirect infallers being as low as 50\percent\ that of direct infallers into the same low-mass clusters at the same times (\Cref{f:preprocessing_smallhost}). Overall, the effects of pre-processing (mass lost as a satellite of a previous host) and post-processing (mass lost within the $z=0$ host) are equally important in shaping the SHSMR of present-day satellites.

The scatter in the satellite SHSMR depends most strongly on when a galaxy first became a satellite, increasing continuously while galaxies are satellites independently of stellar mass (\Cref{f:scatter_evolved}). The stellar mass dependence of intrinsic scatter in the present-day SHSMR is therefore due to the mixing of galaxies which have been satellites for different time spans.

Finally, we quantify the degree to which mass loss can be estimated  observationally by comparing the satellite and central SHSMRs as done in weak lensing analyses. While this comparison provides a biased estimate, the bias is well behaved, with a steep dependence on cluster-centric radius but independent of stellar mass (\Cref{f:massloss}), suggesting it provides a good observational mass loss proxy.

\section*{Acknowledgements}

The code used for the analysis and to generate figures in this paper is publicly available at \url{https://github.com/cristobal-sifon/eagle-satellites}. The \hbt\ catalogue is available upon request to J.\ Han. 

We thank the anonymous referee for their useful suggestions. CS acknowledges support from the Agencia Nacional de Investigaci\'on y Desarrollo (ANID) through FONDECYT Iniciaci\'on grant no.\ 11191125 and through BASAL project FB210003. JH is supported by NSFC (11973032, 11890691), 111 project (No.\ B20019),  the science research grants from the China Manned Space Project (No.\ CMS-CSST-2021-A03) and the Yangyang Development Fund.

This work used the DiRAC@Durham facility managed by the Institute for Computational Cosmology on behalf of the STFC DiRAC HPC Facility (\url{www.dirac.ac.uk}). The equipment was funded by BEIS capital funding via STFC capital grants ST/P002293/1, ST/R002371/1 and ST/S002502/1, Durham University and STFC operations grant ST/R000832/1. DiRAC is part of the National e-Infrastructure.

This work made use of Astropy: a community-developed core Python package and an ecosystem of tools and resources for astronomy \citep[\url{http://www.astropy.org}]{astropy:2013, astropy:2018, astropy:2022}.
This work has benefited from open-source software including Matplotlib \citep[\url{https://matplotlib.org/}]{matplotlib} NumPy \citep[\url{https://numpy.org/}]{numpy}, SciPy \citep[\url{https://scipy.org/}]{scipy}, and Pandas \citep[\url{https://pandas.pydata.org/}]{Reback2020pandas,Mckinney-proc-scipy-2010}. 
Parts of the results in this work make use of the colour maps in the CMasher package \citep[\url{https://cmasher.readthedocs.io/index.html}]{cmasher}.

\bibliographystyle{aa}
\bibliography{bibliography}

\begin{appendix}

\section{Orphan galaxies and disrupted subhaloes}
\label{s:disruption}

Dark matter simulations produce large populations of fully disrupted subhaloes: subhaloes whose mass has fallen below the resolution of the simulation. Consequently, semi-analytic models introduce the concept of ``orphan'' galaxies: fully disrupted dark matter subhaloes whose stellar content remains bound \citep[e.g.][]{Tollet2017,Han2016}. However it is now quite clear that a majority, if not all, of the disruption is numerical rather than physical in origin \citep{vdBoschO2018,vdBoschOHB2018}. Baryons do not seem to alleviate this: \cite{Bahe2019} have shown that subhaloes are disrupted essentially just as often in the Hydrangea simulations (a set of zoom-in simulations run with the same hydrodynamics model as EAGLE), and if a dark matter subhalo is disrupted so is the galaxy. In other words, orphan galaxies essentially do not exist in the Hydrangea simulation. Hydrangea includes clusters with $\mtwo>10^{14}\,\Msun$; by exploring these issues in EAGLE we therefore provide an extension of those results to lower cluster mass here.

\begin{figure}
    \centering
    \includegraphics[width=\linewidth]{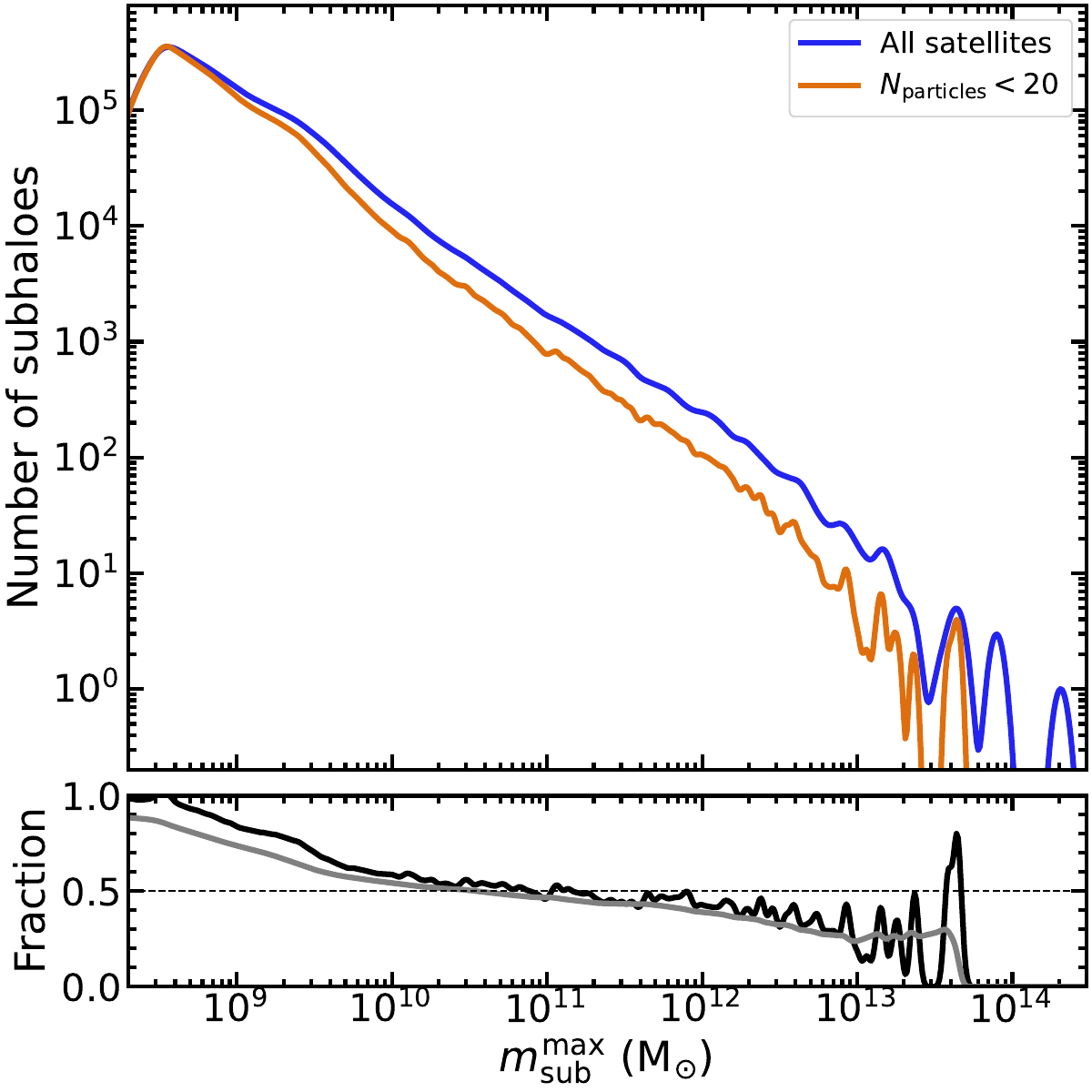}
    \caption{
    Census of disrupted subhaloes. \textit{Top panel:} Unevolved subhalo mass function for present-day satellites in $\mtwo>10^{13}\,\Msun$ clusters. Shown are all subhaloes (blue) and disrupted subhaloes, defined as subhaloes which at present have fewer than 20 particles in total. \textit{Bottom panel:} Fraction of disrupted subhaloes as a function of maximum subhalo mass. The black line shows the differential fraction and the grey line shows the cumulative fraction of disrupted subhaloes above a certain mass. The dashed line marks a fraction of 0.5 for reference. All curves have been smoothed with Gaussian kernel density estimates with $\sigma=0.1$ dex in mass. The lower mass limit approaches the numerical resolution of the simulation.
    }
    \label{f:disruption}
\end{figure}

We show the fraction of disrupted satellites in our cluster sample in \Cref{f:disruption}, as a function of maximum subhalo mass. Here, we define a subhalo as disrupted if it has a total of 20 bound particles or fewer, considering all species. The disrupted fraction increases noticeably with decreasing maximum mass but even subhaloes with maximum masses $>10^{12}\,\Msun$ have a nonnegligible probability of being disrupted. Overall, the fraction of disrupted subhaloes is consistent with the fraction typical of dark matter simulations \citep{Han2016}, despite a lower fraction at the very high mass end which could be dominated by recent infallers.

Consistently with \cite{Bahe2019}, we find little evidence for orphan galaxies. There is only one galaxy in our sample of satellites with fewer than 20 dark matter particles (it has 18) and only seven with less than 100 dark matter particles ($m_\mathrm{DM}<10^9\,\Msun$). Even at lower stellar masses there are hardly any galaxies deficient in dark matter, with only 18 galaxies (out of 28,937) with $\mstar>10^7\,\Msun$ and less than 20 dark matter particles ($m_\mathrm{DM}<2\times10^8\,\Msun$). Only four satellite galaxy (residing in $\mtwo>10^{13}\,\Msun$ clusters) have lost \emph{all} of their dark matter, retaining $\leq1$ dark matter particles. The most massive of these four has a stellar mass $\mstar=5\times10^{7}\,\Msun$ and therefore none of them are included in the analyses in this paper. These results are also consistent with \citet{Jing2019} who found only $\sim 3\%$ of satellites in EAGLE clusters have a dark matter fraction $<50\%$.

Combined, these two results imply that, as already discussed by \cite{Bahe2019}, baryons do little to stop subhalo disruption in simulations. As it is likely that a large fraction of the observed disruption is numerical, there is a population of galaxies in observations which is entirely absent in simulations, both in dark and stellar matter. Contrary to the assumption that numerical disruption in \nbody\ simulations should result in a population of orphan galaxies, to a large extent it appears the dark matter subhaloes should survive too. We do not know whether the properties of those subhaloes are statistically identical to those of surviving subhalo, so this is a significant source of systematic uncertainty in analyses of the satellite populations in numerical simulations.
While a full exploration of these issues is beyond the scope of this work, we end by noting that there appears to be no significant difference in the fraction of disrupted subhaloes as found by \subfind\ \citep{Bahe2019}, \textsc{rockstar} \citep{vdBoschO2018}, or \hbt\ as in this work, so disruption seems to be a feature of the simulations and not the subhalo finders.

\section{Pre-processing and post-processing of satellites in low-mass clusters}

\begin{figure*}
    \centering
    \includegraphics[width=\linewidth]{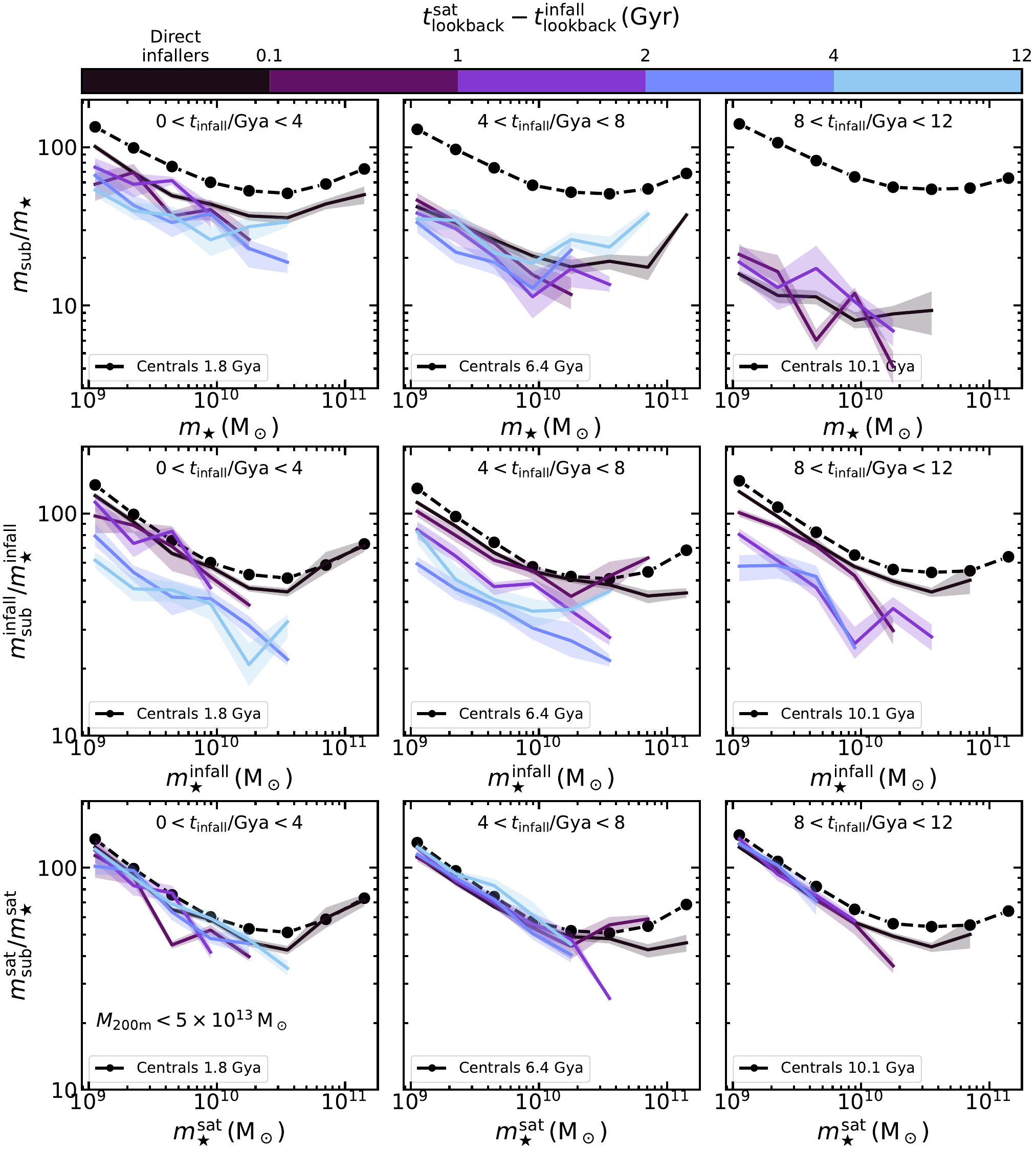}
    \caption{Impact of pre- and post-processing on the satellite SHSMR over time, for satellites currently residing in clusters with $\mtwo<5\times10^{13}\,\Msun$. Binning and styles are as in \Cref{f:hsmr_evolved}.}
    \label{f:preprocessing_smallhost}
\end{figure*}

\begin{figure*}
    \centering
    \includegraphics[width=\linewidth]{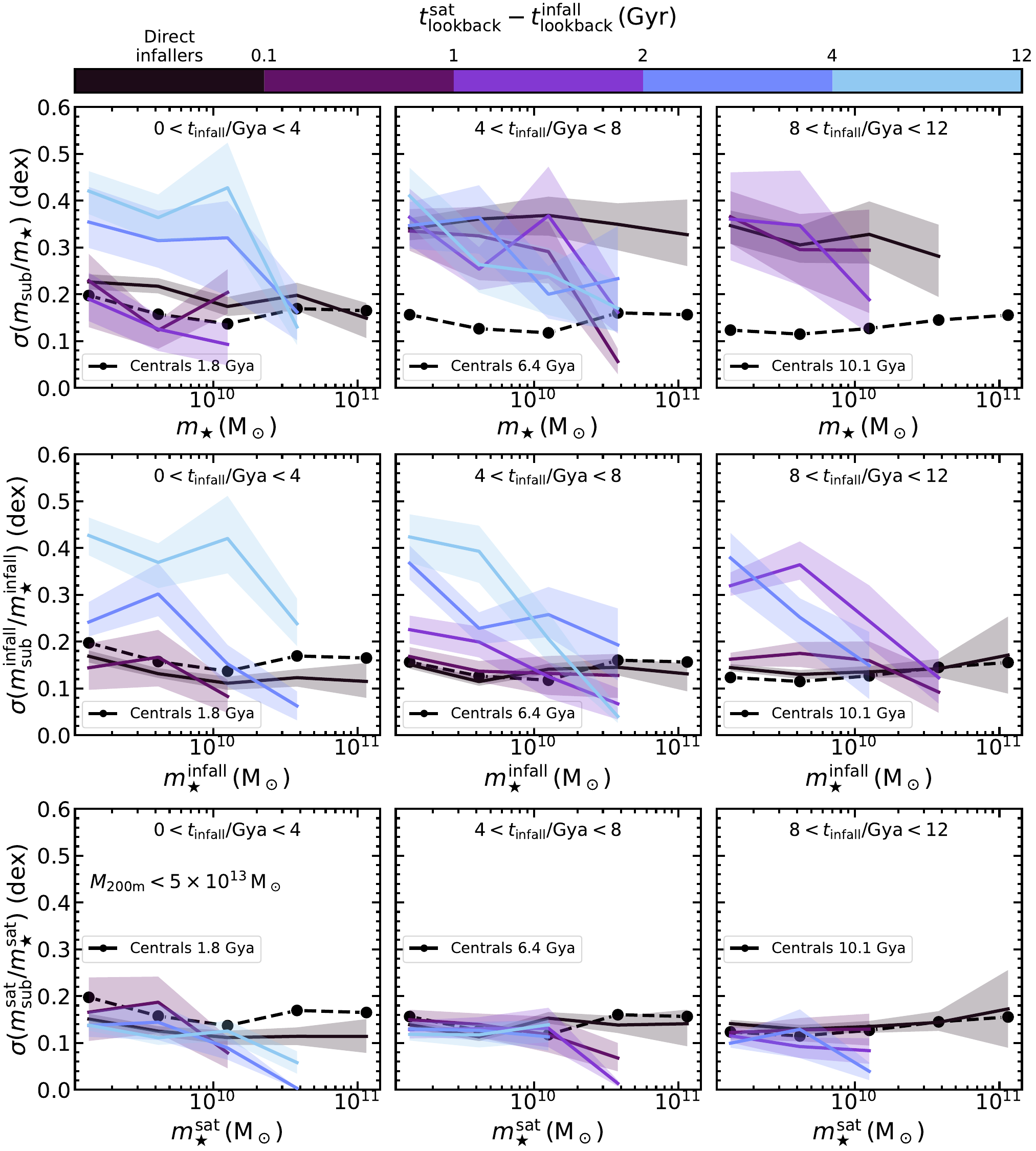}
    \caption{Impact of pre- and post-processing on the satellite SHSMR scatter over time, for satellites currently residing in clusters with $\mtwo<5\times10^{13}\,\Msun$. Samples, binning, and styles are as in \Cref{f:preprocessing_smallhost}.}
    \label{f:preprocessing_smallhost_scatter}
\end{figure*}

Here we show the equivalent of \Cref{f:hsmr_evolved,f:scatter_evolved} but for satellites hosted by low-mass clusters, $10^{13} < \mtwo/\Msun < 5\times10^{13}$. Trends are qualitatively consistent with \Cref{f:hsmr_evolved,f:scatter_evolved} but the fractional amount of mass lost is lower in these satellites at all stages. See \Cref{s:hsmr_evol_mean,s:hsmr_evol_std} for details.

\end{appendix}

\end{document}